\documentclass[letterpaper,twocolumn,10pt]{article}
\pdfoutput=1
\usepackage{usenix2019_v3}

\usepackage{hyperref}
\usepackage{filecontents}
\usepackage{balance}
\usepackage{soul}
\usepackage{listings}

\usepackage{makecell}
\usepackage{amssymb}
\usepackage{pifont}

\usepackage{graphicx}
\usepackage{tabularx}
\usepackage{booktabs}
\usepackage{url}
\usepackage{breakurl}
\usepackage{etoolbox}
\usepackage[numbers]{natbib}
\usepackage{enumitem,amssymb}
\newlist{todolist}{itemize}{2}
\setlist[todolist]{label=$\square$}
\usepackage{pifont}

\usepackage[makeroom]{cancel}

\usepackage{graphicx}
\usepackage{subcaption} 
\usepackage{comment}
\usepackage{xspace}

\usepackage{multirow}
\usepackage{soul}

\usepackage{amsthm}
\usepackage{amsfonts}
\usepackage{mathtools}
\usepackage[utf8]{inputenc}

\usepackage{color}
\usepackage{xcolor}

\usepackage{enumitem}
\setlist[itemize]{noitemsep, topsep=0pt, leftmargin=10pt}
\setlist[enumerate]{noitemsep, topsep=0pt, leftmargin=10pt}
\usepackage{colortbl}
\definecolor{Gray}{gray}{0.65}
\definecolor{LightGray}{gray}{0.9}

\usepackage{array}
\usepackage{arydshln}
\newcommand{\todo}[1]{\textcolor{red}{\small{TODO: #1}}}
\newcommand{\todohs}[1]{\footnote{\textcolor{magenta}{\small{H: #1}}}} % Hossein
\newcommand{\blue}[1]{\textcolor{blue}{#1}}

\newcommand{\red}[1]{\textcolor{red}{#1}}

\newcommand{\shorten}[1]{}

\newcommand{\lsec}[1]{\label{sec:#1}}
\newcommand{\lfig}[1]{\label{fig:#1}}
\newcommand{\ltab}[1]{\label{tab:#1}}
\newcommand{\lequ}[1]{\label{eq:#1}}
\newcommand{\rsec}[1]{\S\ref{sec:#1}}
\newcommand{\rfig}[1]{Fig.~\ref{fig:#1}}
\newcommand{\rtab}[1]{Table~\ref{tab:#1}}
\newcommand{\requ}[1]{Eq.~\ref{eq:#1}}

\newcommand{\secspacingtop}{\vspace{-7pt}}
\newcommand{\secspacingbot}{\vspace{-5pt}}
\newcommand{\subsecspacingtop}{\vspace{-5pt}}
\newcommand{\subsecspacingbot}{\vspace{-2pt}}

\newcommand{\oursystem}{TimeCrypt\xspace}
\newcommand{\oursystemPlus}{TimeCrypt+\xspace}
\newcommand{\mac}{HoMAC\xspace}
\newcommand{\macs}{HoMACs\xspace}

\newcommand{\chunksum}{digest\xspace}
\newcommand{\chunksums}{digests\xspace}
\newcommand{\enc}{HEAC\xspace}

\newcommand{\Mod}[1]{\ \mathrm{mod}\ #1}

\newcommand{\setsmallcaption}{\small} 

\newcommand{\fakeparagraph}[1]{\vskip 0pt\noindent\textbf{#1 }}

\begin{document}

%\title{\bf \oursystem: Encrypted Statistical Index for Time Series Data}
%\title{\bf \oursystem: Private Queries over Encrypted Time Series Data}
%\title{\bf \oursystem: A scalable Private Time Series Database}
%\settopmatter{printfolios=true} 
\title{  %A Scalable Private Time Series Data Store \\ OR \\
\Large \bf %TimeCrypt: A Scalable Encrypted Time Series Database with Cryptographic Access Control
TimeCrypt: Encrypted Data Stream Processing at Scale with Cryptographic Access Control \vspace{-1em}
}
%Access-Policy Driven Processing of Encrypted Data Streams at Scale}
%\title{\bf \oursystem: Processing Encrypted Time-Series Data\\ with built-in Access Control}

\author{Paper \#204, 12 pages + references 
}

\patchcmd{\maketitle}
	{\@maketitle}
	{\vspace{-3em}\@maketitle\vspace{-.5em}}% change the value as needed
	{}
	{}
\makeatother

%\author{Lukas Burkhalter}
%\affiliation{\institution{Department of Computer Science\\ ETH Zurich, Switzerland}}
%\email{lubu@inf.ethz.ch}
%
%\author{Hossein Shafagh}
%\affiliation{\institution{Department of Computer Science\\ ETH Zurich, Switzerland}}
%\email{shafagh@inf.ethz.ch}
%

%\author{Anwar Hithnawi}
%\affiliation{\institution{UC Berkeley}}
%\email{hithnawi@berkeley.edu}

%\author{Sylvia Ratnasamy}
%\affiliation{\institution{UC Berkeley}}
%\email{sylvia@berkeley.edu}

%USENIX for single author (just remove % characters)
%\begin{comment}

\newcommand*{\affmark}[1][*]{\textsuperscript{#1}}

\author{
{\rm Lukas Burkhalter\affmark[1], Anwar Hithnawi\affmark[1,2], Alexander Viand\affmark[1], Hossein Shafagh\affmark[1], 
Sylvia Ratnasamy\affmark[2]}  \\ \\
% \vspace{10pt}
%Department of Computer Science\\ 
\affmark[1]ETH Z{\"u}rich     \hspace{15pt}
\affmark[2]UC Berkeley%, Switzerland} 
} % end author
%%%\end{comment}

\date{}

\maketitle

% Use the following at camera-ready time to suppress page numbers.
% Comment it out when you first submit the paper for review.
%\pagestyle{empty}

\begin{abstract}

%\todo{Why do we need strong protection of sensitive time series yet retain its value by supporting rich functionality and abstractions important without impair the scalability or performance of time series systems ? And what makes this paper novel?} 

A growing number of devices and services collect detailed \textit{time series} data that is stored in the cloud.
Protecting the confidentiality of this vast and continuously generated data is an acute need for many applications in this space.
At the same time, we must preserve the \emph{utility} of this data by enabling authorized services to \emph{securely} and \emph{selectively} access and run analytics. 
This paper presents \oursystem, a system that provides \textit{scalable} and \textit{real-time} analytics over large volumes of encrypted time series data.  
\oursystem allows users to define expressive data access and privacy policies and enforces it cryptographically via  
encryption. 
In \oursystem, data is encrypted end-to-end, and authorized parties can only decrypt and verify \textit{queries} within their authorized access scope.
Our evaluation of \oursystem shows that its memory overhead and performance are competitive and close to
operating on data in the clear.  

\end{abstract}

% !TEX root = ../paper_secure_TS.tex
%\vspace{-10pt}
\secspacingtop
\section{Introduction}
\secspacingbot
\lsec{intro}

Recent years have seen explosive growth in systems and devices that collect time series data and relay it to cloud-based services for analysis.
This growth is only expected to accelerate with the proliferation of IoT devices, telemetry services, and improvements in data analytics.
However, with this growth has come mounting concerns over data protection and data privacy~\cite{iot-leakage-imc}.
%Public's concern over data privacy and confidentiality are unprecedented today in the light of numerous data breaches
Today, the public concern over data privacy and confidentiality is reaching new heights in light of the growing scale and scope of data breaches
~\cite{techcrunch, capital-one, enterprise-data-breach}. 
To grasp the extent of this issue, one can look at the number of data breaches reported under the new GDPR \emph{obligation to notify}, which has already exceeded 65,000 in the first year~\cite{gdpr}.
%In the first year since the European General Data Protection Regulation (GDRP) came into effect alone, more than 65,000 data breaches were registered~\cite{gdpr}.
%Common causes of data breaches are lack of proper data protection (e.g., no fine-grained access control~\cite{VijayGDPRprivacy}),
%malicious insiders (privilege misuse) or insider errors (e.g., misconfiguration or lack of secure configuration), and third-party data access.}

%- Physical access to the bus, and vulns to root the host
%- Privileged admins or insiders 
%- misconfiguration of security properties (Capital one)
%- Bugs in the OS/hypervisor
%- Third parties accessing data without your consent
% of breaches, insider and outsider threats, or unintentional data leaks (e.g., due to bugs or misconfigurations).}

Over the last decade, encrypted databases~\cite{cryptdb, monomi, talos, seabed, blindseer} have emerged as a promising solution to tackle the problem of data breaches.
The approach of keeping data encrypted while in-use allows users to query encrypted data while preserving both confidentiality and functionality.
Research in this domain
has led to various encrypted database designs, including designs for key-value stores~\cite{VLD}, batch analytics~\cite{seabed}, graph databases~\cite{g1}, and relational databases~\cite{cryptdb, monomi}.
This motivates the following natural question: 
\emph{can we enable encrypted data processing for time series workloads?} \newline
Time series workloads come with unique performance and security requirements that  existing encrypted data processing systems fail to meet:

\fakeparagraph{(i)~Scalability and Interactivity.} 
Query processing over time series data must simultaneously scale to large volumes of data, support low-latency interactive queries, \emph{and} sustain high write throughput. 
To meet these challenges several dedicated databases have been designed for time series workloads~\cite{gorilla, BTrDB, influxdata, netflix, kairosdb, timescale, chronix}. 
A key aspect of these systems is their use of in-memory indices that store aggregate statistics, enabling faster query response times and data summarization. 
%\todo{one sen tence that we need to sustain this requirements. Maybe more on highthrouput writes. Data is unbounded and continues. }
As we discuss in \rsec{evaluation}/\rsec{related-work}, prior work on encrypted data processing does not easily lend itself to 
maintaining these in-memory indices. 
The overhead of the crypto primitives in encrypted data processing needs to be negligible to meet the scaling, latency, and performance requirements associated with time series workloads.

\fakeparagraph{(ii)~Secure Sharing.}
A key challenge in modern systems is that privacy must co-exist with the desire to extract value from the data, which typically 
implies \emph{sharing} data to be analyzed by third-party services~\cite{databox}.
Hence, a truly comprehensive approach to data protection must also comprise mechanisms for secure sharing of encrypted data.
Sharing should also be \emph{fine-grained} since it is undesirable and often unnecessary to give parties unfettered access to the data. 
Instead, users may want to 
\textit{(1)}~share only aggregated statistics about the data (e.g., avg/min/max), 
\textit{(2)}~limit the resolution at which such statistics are reported (e.g., hourly vs. per-minute),
\textit{(3)}~limit the time interval over which queries are 
issued (e.g., only June 2019), 
\textit{(4)}~or a combination of the above. Moreover, the desired granularity and scope of sharing can vary greatly across users and applications.
% \remove{even authorized parties the ability to query the data in an unfettered manner.}
Hence, support for encrypted query processing must go hand-in-hand with access control that limits the scope of data that users might query.
The sharing paradigm in data-stream systems is distinctly different than in conventional databases.
Data-stream settings feature a multitude of data sources continuously pushing data to the cloud, where various services 
that are often not known in advance can subscribe to consume and analyze data streams.
Therefore, such systems require flexible access policies.
%\blue{There have been efforts to build private storages such as databox~\cite{databox} that provide fine-grained access control for data providers, but they require storing data locally and do not operate on encrypted data.} %\todo{I would remove this sentence.}
%\blue{Since } services are often not known in advance \blue{we need to} support flexible access polices. 
Frequently, there is a need to fuse and analyze data from different sources collectively; 
this implies that we need to devise an end-to-end encryption scheme that is compatible with this sharing 
paradigm.

%%%%%%%%%%%%%%%%%%%%%%%%%%%%%%%%%%%%%%%%%%%%%%%%%%%%%%%%%%%%%%%%%%%%%%%

\fakeparagraph{\oursystem.}
In this paper, we present \oursystem, a system that augments time series databases with efficient encrypted  
data processing.
\oursystem provides cryptographic means to restrict the query scope based on data owners' defined policies.
%we address the challenges of building a query processing system over encrypted time series data.
%We present \oursystem, a system that augments time series databases with efficient and privacy-preserving 
%processing of time series data, while providing cryptographic means to restrict the query scope based on data owners policies.
%Particularly, we introduce a new \textit{encryption-based access control}  construction (\enc) that is \textit{additively homomorphic}. 
%\todoav{Language issue: grammatically speaking you just said you augment time series DB with timeseries processing}
%\todo{Yes, but WHY? Maybe add a short clause to this sentence: "to provide privacy-preserving <something>".}
%\todo{Why should I care about your new constructions homomorphic-ness at this point? Instead, first tell me what the system can do for me: \oursystem provides scaleable cryptographic access control that supports important time-series concepts like ranges and granularity. THEN tell me how you did it: To achieve this, we introduce a new ... constructions that is add. hom.}
With \oursystem, data owners can cryptographically restrict user A to  
query encrypted data at a defined temporal range and granularity,
while simultaneously allowing user B to execute queries on the same data at a different granularity without
\textit{(i)}~introducing ciphertext expansion or data redundancy,
\textit{(ii)}~introducing any noticeable delays, or 
\textit{(iii)}~requiring a trusted entity to facilitate this.
%\todoah{(1) a new approach to encrypting stream data that is both support both secure computation and access control at the same time. 
%(2) notion of encoding to support rich queries from a scheme that is only additive. 
%(3) access control construction core idea. fine-grained and flexible yet scalable.}

%\red{\fakeparagraph{\enc.}}
In this work, we introduce a partially homomorphic-encryption-based access control construction (\enc) that supports both fine-grained 
\emph{access control} and \emph{computations} over encrypted data within a \emph{unified scheme.} 
These two aspects have traditionally been addressed independently:
the former through cryptographically enforced access control schemes~\cite{abe5, abe2, goyal2006attribute, jedi, Sieve}
and the latter through encrypted data processing~\cite{cryptdb, monomi, talos, seabed}.
\enc simultaneously supports both while meeting the performance and access control requirements of time series workloads.
%The encryption scheme in \oursystem should adhere to the many-to-many sharing paradigm, where sources encrypt their data streams, and data consumers, unknown in advance, 
%can access data according to varying granular access polices defined by the data owner.
%This requires that the data stream is encrypted with frequent key rotation, such that it is feasible to express flexible policies covering encryption keys which correspond to the shared data stream segment.
%More importantly, such scheme should support computation on data that is encrypted with different keys.
%Homomorphic stream ciphers come to mind as the natural approach.
%However, they require overcoming several inherent challenges, namely:
%managing the massive state of keys and identifiers, mapping of keys to access and privacy policies, and also providing a mean to verify the correctness of the outsourced computation.
%\todoav{Language: The last thing wasn't challenges, but a list of three cool properties. Also, it feels weird to start a section\fakeparagraph with a "these" that refers to something outside the paragraph} 
%\shorten{To realize the properties of \oursystem, we devise a novel additively Homomorphic Encryption-based Access Control scheme (\enc) for stream data.}
%\todo{This section does not mention computation nor the key canceling idea, we should be more upfront with the controbution here:}
A key insight behind the design of \enc is based on the observation that time series data streams are continuous and time is the natural attribute for accessing and processing this data.
Hence, we discretize data streams into fixed-length time segments, and encrypt each segment with a different key using symmetric-key homomorphic encryption.
This allows us to express fine-grained access policies at the stream segment granularity. 
This, however, raises two challenges; 
we need to manage a large number of keys in an efficient and scalable manner and translate stream access policies to the corresponding keys succinctly.
To overcome these challenges in \enc, we associate keys with temporal segments.
With this, we avoid the need to maintain a mapping between keys and ciphertexts.
We derive these keys from a hierarchical key-derivation tree construction that allows us to express fine-grained access policies over stream data and share keys efficiently (i.e., with logarithmic complexity).

We provide an implementation and evaluation of a prototype of \oursystem on top of Cassandra.
We evaluate \oursystem in a range of scenarios combining IoT devices, AWS (for data storage and processing), and time series traces from real-world applications. We show that \oursystem can support a wide range of applications by
developing four applications which vary in complexity and scalability requirements.
Finally, we show that \oursystem's performance is competitive with the baseline (plaintext) and it outperforms prior work by a factor of 2 to 52~(\rsec{evaluation}).
Considering an ingest workload with 5.77~million data points per second \textit{on a single machine}, \oursystem's throughput is 
reduced only by 2.9\% for both data ingest and statistical queries over encrypted data. \vfill\eject

\fakeparagraph{Contributions.}
In summary, our contributions are:
\begin{itemize}
\item We introduce HEAC, an encryption-based access control construction for stream data that is additively homomorphic. 
	HEAC additionally provides verifiable computations over ciphertexts to ensure the integrity of the outsourced encrypted computation.
\item We design, implement, and evaluate \oursystem, the first scalable privacy-preserving time series database that meets the scalability and low-latency requirements associated with time series workloads.
We introduce a design that protects the data confidentiality, yet maintains its utility by efficiently supporting a rich set of functionalities and analytics
that are key to time series data.
%\red{\oursystem also supports queries that span multiple streams, e.g., from multiple users.
%\todo{How? In 3.4 we have that it's because "combining ciphertexts from multiple streams creates valid ciphertexts under a new, virtual, shared key"  (already slightly adjusted for the setting)}
%Accordingly, a client can decrypt a multi-stream query result only if she is granted access to all streams %involved in the inter-stream processing.}
\oursystem supports data lifecycle operations such as ongoing data summarization and deletion that are common in time series databases. \oursystem supports expressive data access and privacy policies, enforceable by encryption.
\item We make \oursystem's code publicly available\footnote{Available at: \url{https://timecrypt.io/}}, both as a standalone system and as a library to be integrated with other time series databases.
\end{itemize}

\secspacingtop
\section{Overview}
\secspacingbot
\lsec{overview}
%\vspace{-3pt}

\oursystem achieves its competitive performance through a careful design of cryptographic primitives tailored for time series data workloads.
To understand the rationale behind our techniques, we start this section by presenting relevant background on time series data, then we give an overview of \oursystem, and describe our security model.
%\todo{add paragraph about access revocation in the design.}

%We now discuss the key challenges we tackle in this work,
%give an overview of \oursystem, and describe our security model.
%Afterward, we describe a strawman construction, highlight its shortcomings, and make a case for \oursystem.

%\todo{For time series data, the datastore should handle heavy write workloads, such as Cassandra.
%The advantages of in-memory index are
%\textit{(i)}~faster response time for processing statistical queries compared to fetching data from database and computing on the fly.
%\textit{(ii)}~reducing the load and number of severs needed to handle the requests.
%\textit{(ii)}~increased throughput.
%}

%\vspace{-5pt}
%\subsection{Challenges}
%\vspace{-5pt}
%\lsec{strawman}

\shorten{We start this section by discussing three key requirements of time series applications that were not supported by prior 
encrypted data management systems, which led us to design, a new system: \oursystem. 
Afterwards, we present an overview of \oursystem and discuss our threat model.}

\subsecspacingtop
\subsection{Background on Time Series Data}
\subsecspacingbot
\lsec{bg:timeseries}
  %\vspace{-5pt}
  
\fakeparagraph{Time series Applications.} 
Time series data is increasingly prevalent across a wide range of systems (e.g., monitoring,  telemetry, IoT) 
in diverse domains such as health, agriculture~\cite{farmbeats-MSR}, transportation~\cite{car-data}, operational 
insight~\cite{dev-ops-applications}, and smart cities.
The growth of time series data is largely attributed to the rising demand for instrumentation. 
Individuals and organizations are continuously logging various metrics which report the state of systems or organisms for better 
diagnoses, forecasting, decision making, and resource allocation. 
The ability to capture and analyze this data in a timely manner is key for automation and is enabling a whole new spectrum of 
applications~\cite{farmbeats-MSR, car-data,netflix,dev-ops-applications,bolt}.
The proliferation of time series data has been coupled with increasing demand for high-performance analytics over large volumes 
of time series, and has 
led to numerous designs for databases that are optimized for time series workloads~\cite{gorilla, BTrDB, influxdata, netflix, kairosdb, timescale, chronix}.

\fakeparagraph{Time series Workloads.} 
%The distinguishing characteristics of  time series analytics are:
\textit{(i)~Write and Read:} 
Data is append-only and typically generated at an extremely high rate (high velocity) and is initially stored at a 
high resolution (large volume)~\cite{gorilla, BTrDB}.  
It is not unusual for applications in this space to report hundreds of millions of data points per 
day~\cite{BTrDB,summarystore}. Hence, sustaining high read and write throughputs and scalability are key requirements when 
storing and processing time series data.
Time is the primary dimension for accessing and processing data.
Queries  primarily consider temporal ranges 
%(e.g., return all values from Jan 25 - Feb 25) 
(e.g., values from the last 3h)
rather than targeting individual points.
%-- i.e., by specifying a contiguous time interval rather than target point queries. 
%where data is accessed by specifying an arbitrary time interval, i.e., a set of temporally co-located data points rather than target point queries.
%Time is a primary axis for accessing and processing data, 
%as captures how measurements change over time.  
%and mostly at a lower resolution, e.g., minute aggregates.
\textit{(ii)~Analytics:}
Queries are primarily of aggregate and statistical nature,
%To keep computation tractable over massive time series data, 
and specialized indices for accelerating statistical queries are common in time series databases~\cite{BTrDB, influxdata, 
prometheus}. 
Additionally, analytics of diagnostic (e.g., anomaly or trend detection) and predictive nature (e.g., forecasting) are common in this 
space.
\textit{(iii)~Data decay:} 
Time series data is often machine generated, continuous, and massive.
Simultaneously, the value and relevance of data decays rapidly with time.
Analytics largely favor recent data over older, and roll up aggregation is commonly applied to older data to reduce storage requirements. 
Hence, data retention and summarization~\cite{summarystore,gorilla} are crucial for these systems. 

The goal of our work is to retain the performance, functionality, and scalability of existing time series databases while augmenting them with strong security and privacy guarantees.

%\fakeparagraph{Time-series Databases.}\todohs{remove section, content redundant, merge citations?}
%The increase in demand for high performance analytics over large volumes of time-series, has 
%led to numerous designs for databases that are optimized for time-series workloads~\cite{gorilla, BTrDB, influxdata, netflix, kairosdb, timescale, chronix}.
%%\footnote{Time series databases have been the fastest growing category of databases in the last two years with over 50 new emerging systems~\cite{db-rank,TSList}, and with major Cloud providers~\cite{azure, google1, aws1, aws2} offering dedicated systems to process and store large-scale time series data.}. %opentsdb
%Such optimizations include employing simplified data models~\cite{gorilla} and
%building specialized in-memory indices~\cite{BTrDB, Faloutsos, gorilla} or precomputed statistical aggregates~\cite{BTrDB, summarystore,influxdata} to accelerate queries over high volumes of data. 

%Hence, allow for fast stream processing routines. 
%that can quickly identify fractions of the data (often less than 6\%~\cite{bailis2017macrobase}) that are of interest for more complex analytics.

\subsecspacingtop
\subsection{Architecture}
\subsecspacingbot

\global\csname @topnum\endcsname 0
\begin{figure}[t]
	\center
	\vspace{7pt}
	\includegraphics[width=.97\columnwidth]{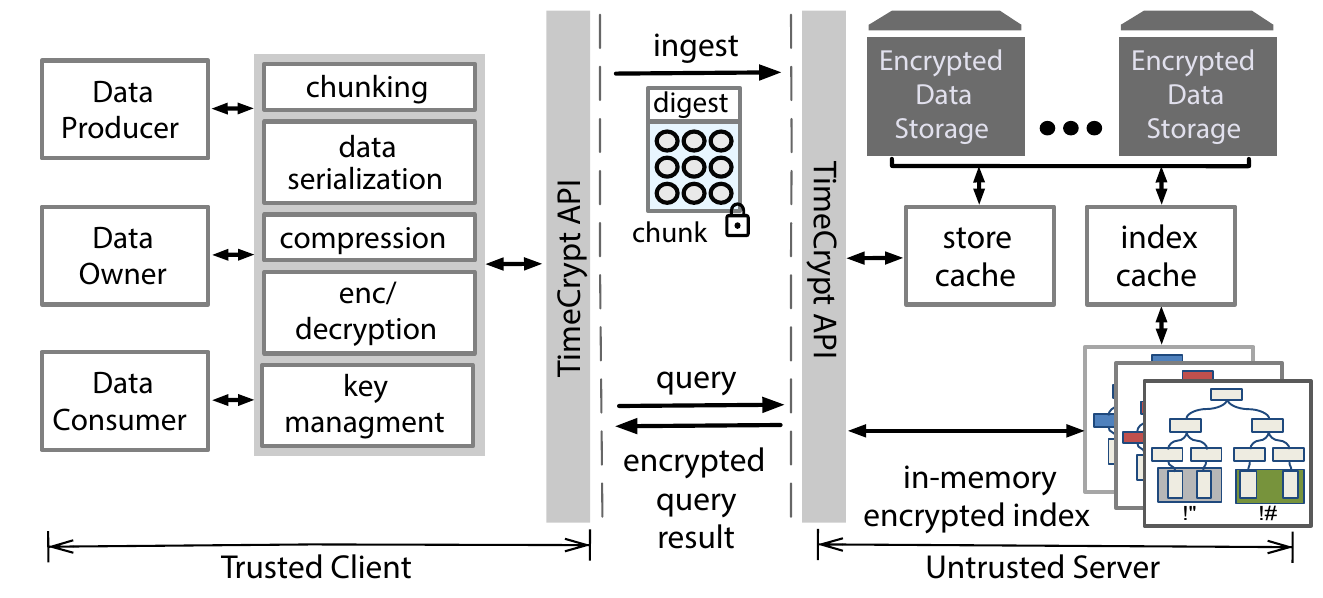}
	\vspace{-10pt}
	\caption{
		\setsmallcaption
		\oursystem's architecture.
		%with a pluggable distributed key-value store.
		%	The pluggable storage requires an efficient, scalable key-value store.
		%Queries hit \oursystem instances, which update the indices on data ingest and process queries.}
		%	Since queries and inserts are focused on the latest time interval, the index is typically in cached (warm queries).
		%	In case the index is not in cache (cold queries), it is fetched from the storage.
	}
	\vspace{-15pt}
	\lfig{arch}
\end{figure} 
%We design \oursystem so its amendable to adoption. 
\oursystem's architecture is analogous to that of conventional times-series databases~\cite{BTrDB, chronix, influxdata,aws1,aws2}, 
where a standard distributed key-value store is extended with additional logic for time series workloads.
\oursystem includes a trusted \emph{client library} to realize end-to-end encryption paired with access control and integrity verification.
%\red{Raw data access similar to other plaintext time series databases~\cite{BTrDB,pisa,influxdata}.}
%The server also computes a succinct tag as the computation proof on encrypted result.
%\todo{One reviewer got confused why we differentiate between data producer and data owner. We should add one statement on owner and producer encryption relation similar to that in Droplet. Or otherwise simplify and have both called as owner.}
\oursystem consists of two components (i.e., the client and server libraries) and
involves four parties (i.e., data owner, data producer, data consumer, and database server), as illustrated in \rfig{arch}.
%Data producers  %, \remove{such as IoT devices.}
A \textit{data producer} is an entity (e.g., IoT device) that generates and uploads time series data, and runs \oursystem's client library which handles stream preprocessing and encryption. 
%securely storing the data on the server.
The \emph{data owner} % of the generated data %are entities who own the data generated by data producing devices and 
can express access permissions to its generated data. %~(\rsec{implementation}). 
Meanwhile, \textit{data consumers} are entities (e.g., services) that are authorized to access a user's data
to provide added value, such as visualizations, monitoring, and diagnoses.
%Data consumers leverage \oursystem's client engine to issue queries~(\rsec{functions}).
%The server engine augments the database with encrypted data processing.
\oursystem's server executes statistical and analytical queries directly on encrypted data.
\oursystem supports a rich set of foundational queries that are widely used in time series workloads~(\rsec{functions}), i.e., statistical queries (e.g., min/max/mean), analytics (e.g., prediction, trend detection), and lifecycle operations (e.g., ongoing data summarization, deletion).
%, allowing applications to perform complex computations.
The server builds \emph{in-memory encrypted indices} to support fast queries and analytics~(\rsec{index}).
\shorten{Our encryption-based access control construction enables fast data processing and flexible and fine-grained access control, while maintaining low query response time~(\rsec{efficient-dec}).}

\subsecspacingtop
\subsection{Goals for Stream Data Access Control}
\subsecspacingbot
\lsec{ac-requirements}
Encryption is an effective tool for protecting data from external threats, breaches, or malicious providers.
However, a truly comprehensive approach to data protection must also include mechanisms for enforcing access control policies,
to support the privacy and security principles of least privilege and data minimization, 
where data is protected by limiting unnecessary exposure.
%Having the means to control what authorized principals can access based on the contents of the data item; 
%protecting data by limiting unnecessary data exposure.
%Hence, conforming with the privacy and security principles of the least privilege and data minimization.
%
State-of-the-art relational databases have security mechanisms designed for this purpose. The most adopted approach to support access control is based on views\footnote{A view; also referred to as virtual table, is a dynamic window of a subset of the rows and columns in a database.} and row-level access policies. However, specifying effective access control policies necessitates taking into consideration the semantics of data.
%\blue{Views are used to only .}
%\remove{The most adopted approach to support access control is based on  and row-level access policies.}
%\remove{However, specifying effective access control policies necessitates taking into consideration the semantics of data.}
%Hence, the question is how to specify effective access policies for stream data or in other words what constitutes as a \emph{view} in time series databases?
Therefore, we investigated the major state-of-the-art time series 
databases~\cite{influxdata, opentsdb, timescale, prometheus, graphite, BTrDB, amazon-kinensis-access} to understand the state of affairs in stream data access policies. 
We found that the only access policy restriction provided at the database interface, if any, is at the stream unit (i.e., grant or decline access to the entire stream). 
This binary protection level is however too coarse. 
%\todo{rewrite the text above. This should emphasize why binary protection is not enough and how alternative policies can provide better protection. Or I would just remove the above text since we pose the question below of exploring what type of policies are suitable.}
This prompted the following question: What type of policies can offer the fine-grained protection that is required for selective and secure sharing of data streams?
%\todo{We should give an example why fine-grained access control is necessary here}
Stream data access control literature~\cite{carminati-access-on-streams, sandeep-access-control} and time series applications designed for multi-user settings\cite{thingsboard-iot-management, bolt} both 
recognize that policies which are expressed in time, resolution, and attributes are ideal for fine-grained access restrictions on streams.
Examples of such policies can be a user choosing to simultaneously share hourly averages of their measured heart rate with their doctor and per-minute averages with their trainer but only for the duration of their workout session. 
Similarly, a datacenter operator might share resource utilization levels with a tenant but only for the duration of her job.
Our goal is to translate these stream-specific sharing semantics into a cryptographically enforceable access control mechanism.

\subsecspacingtop
\subsection{Threat Model}
\subsecspacingbot
\lsec{sec-model}

Our goal is to maintain the confidentiality and integrity of
computations running on a cloud infrastructure that is potentially subject to 
an adversary that can read and tamper with data and manipulate query execution.
%\remove{We view security from the point of view of a client  which we assume executes the protocol correctly.}
%\blue{We assume that the client-side application implementation to follows the protocol correctly.}
In order to support sharing, we require a public-key infrastructure, such that entities can be identified and that a private/public key-pair can be associated with them.
\oursystem provides the following guarantees in this setting:
%\todo{Include PKI assumption.}

\fakeparagraph{Confidentiality.} 
Data is encrypted using semantically secure encryption before it leaves the client device.
%\remove{Clients store only encrypted data in the cloud, and }
Since decryption keys are never disclosed to the cloud provider,
%\remove{Only encrypted data is accessible within the cloud's processing pipelines. This ensures } 
data confidentiality is guaranteed even in the case of a system compromise or malicious provider.
%A policy restricts access both on a time range and a minimal query resolution in this time range.
Note that we do not employ property-revealing encryption, avoiding their inherent information leakage issues~\cite{Naveed}. 
Our cryptographic access control mechanism ensures that
data consumers can only query and access data according to the access policies defined by the data owner. 
%\blue{Users can grant only the necessary access to their data, i.e., adhere to the security principle of least privilege. 
%Access policies are enforced by encryption, i.e., data is not exposed to trusted entities or intermediaries for access control enforcement.}
%After granting data access, we cannot control what data consumers do with the data thereafter.
%If an adversary compromises a data consumer, the adversary gains access to only the data the consumer is authorized to access
%and is not able to read data beyond what is allowed by the consumer's access policy.
%\remove{Leakage of data encryption keys results only in the disclosure of the data stream segment associated with the disclosed keys.}
%\todo{\remove{What about the multi interval leakage?}}
%\todo{cover  indistinguishability IND-CPA and access control.}
%Confidentiality holds even when server and consumers collude.
%We refer to \rsec{keymanagement} for more details on the semantics of access control in \oursystem.

\fakeparagraph{Integrity.}
\oursystem's integrity protection guarantees that, if a query completes, its output is equivalent to a correct execution on a trusted platform.
Therefore, a malicious server cannot affect the computation, except by denying service. %\remove{by choosing not to execute the query or hold back resources}. 
%Integrity for a stream is guaranteed as long as no data consumer (that has the key for that stream) colludes with the server.
Note that even in case an integrity key for a stream is leaked, confidentiality remains intact.

\fakeparagraph{Access Patterns.}
%\todo{In TimeCrypt, we do not hide the timestamps of time-series data. }
Similar to previous work~\cite{cryptdb, monomi, seabed, pilatus}, \oursystem is non-oblivious, i.e., it does not protect against 
access pattern-based inferences in a trade-off for performance and scalability.  
Therefore, an adversary can learn which data the consumers are authorized to access by observing access patterns.
%(both in memory (i.e., traversal pattern of the encrypted index) and over the network) 
%\todo{If we say this, why are we not doing this? I think this would be a performance killer (not practical).}
\oursystem could be complemented with Oblivious RAM approaches~\cite{Elaine:RAM} to hide these access patterns.
%\remove{We assume that the client-side application implementation to follows the protocol correctly; i.e., it does not intentionally leak secrets that can compromise the security of TimeCrypt.
%Denial-of-service and side-channels are outside the scope of this paper.}
\fakeparagraph{Access Control Collusion.} 
Resolution based sharing in combination with interval sharing is not collusion resistant, even when considering a plaintext system. 
For example, any entity with access to aggregation over the intervals $[t_0, t_2)$ and $[t_1, t_2)$, can trivially derive the aggregation $[t_0, t_1)$ over the overlapped range by computing the difference.
Hence, clients must be careful when sharing different resolutions over overlapping intervals.
Furthermore, \oursystem comes with a trade-off between performance and collusion resistance when sharing non-continuous intervals. 
In the default mode, an adversary with access to two non-continuous intervals can compute the aggregation between the two intervals. 
For cases where this poses a privacy risk to applications, \oursystem provides a mechanism to prevent such collusion (\rsec{crypto:primitives}) at the cost of increased decryption time.
 %\todo{I'm okay with this for now but want to revise the intro of this text. I think it's  a bit ambiguous.}
%\blue{Care must be taken when selecting intervals to share. 
%	For example, any entity with access to aggregations of two different but overlapping ranges can trivially derive the aggregation also over the overlapped range by computing the difference, gaining access to higher-granularity data.
%	Therefore, data owners might want to avoid sharing overlapping ranges with data consumers.
%	If we assume that different data consumers might collude, overlaps must be avoided not just for each consumer individually but across all shared intervals for a given stream.
%	Whether or not this a relevant threat depends on the application scenario.
%}\todoav{my writing here is too stilted/my sentences way too long. Also add leakage here? Note: Remember to keep framing it as a trade-off, not as a weakness}

A full security analysis with formal definitions and  proof sketch can be found in the appendix~(\rsec{appendix}).%of our homomorphic access control construction is available in \rsec{proof}.
%We formalize the security guarantees of our homomorphic access control construction in \rsec{proof}.

%\red{(with bandwidth overhead in $O(log(n))$~\cite{Elaine:RAM}).} \todohs{remove?}

%\vspace{-3pt}
\subsecspacingtop
\subsection{\oursystem Approach}
\subsecspacingbot
%\vspace{-5pt}
%In this section, we present a high-level overview over our system.
\oursystem is a new encrypted time series database design that meets the scalability and low-latency requirements associated with time series workloads. 
We propose a new approach for data stream encryption that supports processing over encrypted data streams, 
computation integrity, and powerful access control within a unified scheme.

\fakeparagraph{Data Abstraction.}
\lsec{stream-writing}
\oursystem stores data points in a stream as time-ordered chunks of predefined time intervals, i.e., $[t_i, t_{i+1})$ with a fixed interval size $\Delta = t_{i+1} - t_i$.
%Temporal data access is mainly defined at chunk intervals.
Each data chunk also includes an encrypted \chunksum that consists of statistical summaries about the underlying data.
The \chunksum enables \oursystem to compute statistical queries over time ranges efficiently, as we discuss next.
At the client, the chunks are encrypted with standard symmetric encryption while the \chunksums are encrypted with HEAC. 

%\blue{our} encryption-based access control scheme for stream data.
%Data is encrypted at the client-side;
%chunks are encrypted with standard authenticated encryption (i.e., AES-GCM) and digests are encrypted with HEAC.
% (i.e., Homomorphic Encryption-based Access Control).
%Each chunk and digest is encrypted with a new cryptographic key derived from our key-derivation construction (\rsec{keymanagement}).

%The encryption approach we take in this work is distinctly different than that assumed in existing encrypted databases.
%Because fully homomorphic encryption, practical encrypted database instead resort to encrypting the data of different practical encryption schemes to support varying types of queries (e.g., deterministic encryption for equality, additive hom for addition, OPE for order, etc). 
%The tight to observation of the nature of analytics and queries often ran on massive time series data. 
%So instead of using several encyption schemes to support different queri types, 

\fakeparagraph{Aggregatable Digests.}
As \enc is additively homomorphic\footnote{An additive homomorphic encryption scheme supports additions on ciphertexts, such that $decrypt(C_1\oplus C_2) = decrypt(C_1) + decrypt(C_2)$.}, it supports secure aggregation of ciphertexts.
However, to support queries beyond sum, we leverage \emph{aggregatable encoding techniques} that exist in literature to support sophisticated statistical and analytical queries over encrypted data.
At a high level, we introduce a per-chunk digest, which  holds a vector of encoded values $\{x_0 , ..., x_n \}$ that are encrypted with \enc.
To process queries, the server computes the aggregate function on the encrypted encodings across different digests.
With this, we can support statistical queries that are inherently aggregation-based (e.g., sum, mean) or can be transformed to be aggregation-based (e.g., min/max, regression) (\rsec{functions}).
%\blue{For example, to compute an average, the digest has to contain the sum and the count of all underlying values in the chunk. 
%With the aggregated sum and count, the querying entity can efficiently compute averages over different ranges.}
%\blue{ raw data? When do we call something an interval (afaik, we use interval for "time in a single chunk" vs a range (=afaik, we use that for multiple chunks)? In the caption of Figure 5, we use "time ranges (i.e. Interval size)" which seems confusing.
%\blue{
%\fakeparagraph{Continuous Data Streams. }
%\oursystem is generally designed for continuous data streams, where data is continuously written. 
%However, in some applications, data writes are event-based with different frequencies.
%This might leak information both through the irregularity of the data streams or the size of the chunks. 
%To ensure that the server only observes a continuous stream of data, \oursystem offers the option to send chunk and digests continuously.
%Furthermore, \oursystem can reduce the risk of volume leakage by omitting the chunks entirely (i.e., send just the digest) or apply padding to the chunks~\cite{minicrypt}.}
%\todo{this important information but I dont think here is the right place for it. Here we give a gist of the approach and this information is too specific to be placed here. We should integrate it somewhere else. }

\fakeparagraph{Encryption and Access Control.}
%To enable encrypted data processing that natively supports access control,
%we model data streams as units of time epochs, where each time epoch is encrypted with a different encryption key.
%This allows us to express access policies at epoch granularity (referred to as chunk).
%and extend it with an efficient key construction to enable expressing stream access policies.}
%\blue{HEAC is based on symmetric homomorphic encryption where a fresh key is used for each message encryption~\cite{seabed, castelluccia2009}.
%However, \oursystem improves the performance by 2x for time-series workloads compared to other systems~\cite{seabed} by closely coupling keys to time (\rsec{keytimeencoding}) and optimizing for in-range aggregations (\rsec{crypto:primitives}).
%Furthermore, HEAC includes efficient integrity protection (\rsec{integrity}), and integrates fine-grained access control capabilities (\rsec{keymanagement}).}
%Since the encryption scheme is based on a stream cipher, it relies on a keystream.
%A key aspect of our scheme is tied to the observation that time series data streams are continuous.
%Consequently, to enable encrypted data processing that natively supports access control,
A key aspect of our scheme is tied to the observation that time series data streams are continuous. Consequently, to enable encrypted data processing that natively supports access control, we model data streams as a series of time segments, where each segment is encrypted with a different encryption key.
We introduce a time-encoded keystream that maps keys to segments of the data stream, such that when a user restricts access to the data stream, only the corresponding range in the keystream is shared with the data consumer (\rsec{keytimeencoding}).
Based on the access policy, a data consumer is provided with the necessary decryption keys via an access tokens.
Access tokens are encrypted with the data consumer's public key (hybrid encryption) and stored at the server.
To enable sharing without enumerating all the keys and to support a \emph{succinct key state}, we derive keys from a hierarchical tree key-derivation construction (\rsec{keymanagement}). 
We also introduce a technique to support restricting access to a particular resolution level~(\rsec{access-resolution}), e.g.,
%To grant a user access to only a defined temporal \emph{resolution}, 
aggregated values at 10-minute resolution.
\secspacingtop
\section{Encryption in \oursystem}
\secspacingbot
\lsec{design}

In this section, we introduce the cryptographic components of \oursystem and present HEAC in more detail.
HEAC, in essence is based on a symmetric homomorphic encryption~\cite{castelluccia2009}.
However, we improve its performance by a factor of 2x for time series workloads by mapping keys to time and optimizing it for in-range ciphertext aggregations.
Furthermore, we extend it to support fine-grained cryptographic access control capabilities tailored to time series data. 
%with an efficient key derivation construction.
%We achieve this by coupling encryption with a novel key derivation construction.
Finally, we ensure computation integrity on encrypted data via Homomorphic Message Authentication Codes.

\shorten{While \macs have previously been introduced %as cryptographic building blocks 
in the literature~\cite{catalano2013practical}, our construction is 
the first to adapt it efficiently for integrity of encrypted data processing}

\subsecspacingtop
\subsection{Symmetric Homomorphic Encryption.}
\subsecspacingbot
\lsec{crypto:primitives}
\shorten{HEAC is in essence similar to a stream cipher, where one-time keys are combined with the plaintext data block for encryption. 
As in traditional stream ciphers, the scheme makes the assumption that the client is in possession of a pseudorandom keystream $\{k_0, k_1, k_2, ...\}$ for encryption and decryption.}
We encrypt an integer $m_i$ from the message space $[0, M-1]$ as
$c_i = Enc_{k_i}(m_i) = m_i + k_i~mod~M$,
with key $k_i \in [0, M-1]$.
Given $k_i$, one can decrypt $c_i$ as 
$Dec_{k_i}(c_i) = c_i - k_i~mod~M = m_i$.
This scheme is semantically secure when the keys are pseudorandom and no key is reused~\cite{castelluccia2009}.
%Since we use modular addition as the encryption function, the scheme is also additively homomorphic.
\newline
Given the aggregated secret keys, one can decrypt the aggregated ciphertexts:
\vspace{-3pt}
\begin{equation}
\lequ{key:aggregate}
% OLD : Dec_{ \sum_{i=0}^{n} k_i}( \sum_{i=0}^{n} c_i ) = \sum_{i=0}^{n} m_i + \sum_{i=0}^{n} k_i - \sum_{i=0}^{n} k_i ~ mod~M
\sum_{i=0}^{n} m_i = Dec_{ \sum_{i=0}^{n} k_i}( \sum_{i=0}^{n} c_i ) =  \sum_{i=0}^{n} c_i - \sum_{i=0}^{n} k_i ~ mod~M
\end{equation}
%\shorten{The ciphertext size in the scheme is limited by $M$ and the homomorphic additions are extremely efficient, as they correspond to modular additions.}
We set $M$ to $2^{64}$, to support all integer sizes, without leaking any information about their original size.

\fakeparagraph{Key Canceling.}
\lsec{efficient-dec}
In the above scheme, the local computation to aggregate keys is linear in the number of aggregated ciphertexts, forcing the client to perform the same amount of computations as the server.
%\remove{With this scheme, the client needs to locally perform aggregations of the keys proportional to the number of aggregated ciphertexts.
%This is not ideal, as the client performs equal amount of computations as the server.}
We reduce this linear overhead to a constant, by leveraging the fact that time series data is generally aggregated in-range (i.e., over a contiguous range in time) as discussed in \rsec{bg:timeseries}.
We can therefore employ \emph{key canceling}~\cite{castelluccia2011CancelOut, bonawitz2017practical,seabed, Aggregatable-prf}.
%\remove{This encoding also becomes relevant later when we discuss how we enforce resolution-based access} \red{(\rsec{access-resolution}).}
This technique will also be relevant later, when we discuss integrity and access control (\rsec{integrity}, \rsec{keymanagement}).
%\remove{The idea here is to select the individual encryption keys such that during aggregation the inner keys cancel each other, and only the outer keys at the beginning and end of the time range are needed for decryption.}
To enable this optimization, we choose the individual encryption keys such that the inner keys cancel each other out during aggregation.
We do this by replacing the individual key $k_i$ with a composite key that links subsequent messages:

\vspace{-2pt}
\begin{equation}
Enc_{k'_i}(m_i) = m_i + k'_i~mod~M,~\text{with}~k'_i = k_i - k_{i+1}
\end{equation}
For decryption of an in-range aggregated ciphertext (\requ{key:aggregate}), we now  require only the two boundary keys:
%access to $k_i$ and $k_{i+1}$ is required:
%\vspace{-2pt}
%\begin{equation}
%m_i = Dec_{k'_i}(c_i) = m_i + k'_i - k'_i~mod~M
%\end{equation}
%\vspace{-2pt}
%\begin{equation}
%Dec_{ \sum_{i=0}^{n} k'_i}( \sum_{i=0}^{n} c_i ) = \sum_{i=0}^{n} m_i + \sum_%{i=0}^{n} k'_i - \sum_{i=0}^{n} k'_i ~ mod~M
%\end{equation}
%The resulting aggregated key consists of only two keys as the other keys are canceling out:
\vspace{-3pt}
\begin{equation}
\lequ{simplified:equation}
\begin{split}
\sum_{i=0}^{n} k'_i  = (k_0 - \cancel{k_{1}}) + (\cancel{k_1} - \cancel{k_{2}}) \dots (\cancel{k_n} - k_{n+1})\\ 
%= k_0 - k_{1} + k_1 - k_{2} + k_{2} \dots - k_n + k_n - k_{n+1}\\ 
%= k_0 - k_{n+1}
\end{split}
\end{equation}
%The advantage of this key encoding becomes relevant while decrypting an in-range aggregated ciphertext.
With key canceling, the decryption time in \oursystem is independent of the number of in-range aggregated ciphertexts. 
This scheme remains semantically secure~\cite{castelluccia2009, Aggregatable-prf}; an attacker without access to the keys cannot exploit the canceling property.
However, when given access to keys for two non-continuous intervals, an adversary could learn aggregates about the skipped time between the two intervals.
For example, when given access to $k_0, \ldots, k_5$ and $k_{10}, \ldots, k_{15}$, they could compute $ \sum_{i=5}^{10} m_i  ~ \Mod~M$, given $k_5$ and $k_{10}$.
The ramifications of this issue arise when users share adjacent intervals in the same stream with small gaps. 
\oursystem provides a hybrid key-canceling mechanism that limits this leakage in a trade-off for longer decryption times.
We split the keys into epochs by replacing some $k_i$ with non-canceling skip-keys $k'_i, k''_i$ in $k_{i-1} - k_i$ and $k_i - k_{i+1}$, respectively. 
With this, we can share one interval per epoch without leakage. 
This increases the cost of aggregations over the epoch borders by two key derivations and one addition.
%\red{In \rsec{ac:leakage}, we show how to prevent this by selectively breaking the cancellation property at specific points in the keystream, at the cost of a slight overhead\todoav{go into more detail, and calculate the specific overhead for a reasonable example setting}.}
%, this countermeasure is not enabled by default in our system.}
%\blue{An encoding that cancels out keys during aggregation is not new; it was %introduced by Castelluccia et al.~\cite{castelluccia2011CancelOut} and was %adopted in other systems~\cite{seabed, bonawitz2017practical}.
%We describe the encoding in details here because it is relevant to %understanding how we realize resolution based access restriction~\rsec%{access-resolution}.}
%However, Castelluccia's encryption does not lend itself to access control as we discuss next.
%However, our \blue{encryption} distinguishes itself in that it does not result in ciphertext expansion
%- as we do not require tracking the \blue{indices} of the encryption keys per ciphertext, due to our time-encoded key generation, as explained in \rsec{keytimeencoding},
%and comes with a novel key construction scheme enabling fine-grained sharing.
%\todosr{Also Castelluccia's encryption doesn't lend itself to access control as we discuss next? The above makes our contribution sound more incremental.}

% \vspace{-3pt}
\fakeparagraph{Time-Encoded Keystream.}
\lsec{keytimeencoding}
%\vspace{-3pt}
%\remove{The tree key derivation allows us to restrict access to segments of keys efficiently.}
In \oursystem, access permissions are expressed with temporal ranges, e.g., Sep-14-15:00 till Sep-17-06:00 2019.
Internally, \oursystem  \emph{chunks} data into fixed time segments of size $\Delta$,
which can be set per stream (e.g., 10~s intervals).
In addition to the raw data points, each chunk is augmented with digests that are used for statistical query processing.
Each chunk is encrypted with a fresh key from the keystream, indexed by the time window of the chunk.
Assuming the data stream starts at timestamp $t_0$, the chunk \chunksum $m_i$ for the interval from $t_i$ to $t_{i+1}$ is encrypted as  $c_{i} = Enc_{k_i - k_{i+1}}(m_i)$.
By mapping keys to temporal ranges, a time range implicitly determines the position of the used key in the keystream.
As a result, we sidestep the need to store identifiers of the keys along with the ciphertexts and avoid ciphertext expansion.
\subsecspacingtop
\subsection{Integrity}
\subsecspacingbot
%\vspace{-5pt}
\lsec{integrity}

\shorten{\red{MOTIVATION:} \blue{Existing encrypted data processing systems using homomorphic encryption~\cite{cryptdb, monomi, seabed}  focus on protecting confidentiality of the data, but lack in providing mechanisms to verify the results of aggregation queries. This is a concern, especially for applications which make critical decisions based on these query results. An insider with administrator privileges can influence decisions by simply adding an error term to the result or return an altered value.
Even if the server has no malicious intentions, a failure in the computation has the same effect and is hard to detect without any integrity measures. There are several systems that focus on verification of outsourced computation using zero-knowledge proofs~\cite{pinocchio, braun-verify, zql-verify}, but they assume that the prover (i.e., the server) has access to the plaintext data. The combination of above verification and encrypted data processing techniques would result in several orders of magnitude performance reduction considering the massive time series workloads in this work. 
}}

Homomorphic encryption schemes are by design malleable, and therefore susceptible to ciphertext manipulation.
In our setting, a dishonest server could try to drop, duplicate, or manipulate ciphertexts, resulting in incorrect query outputs.
Incentives for deviations from the protocol could be as simple as trying to preserve resources by reducing the complexity of queries~\cite{vc3}.
%The incentive \remove{to not adhere to the protocol can} \blue{could} be as simple as a dishonest cloud provider being tempted to reduce the complexity of queries to preserve resources~\cite{vc3}.
%(i.e., return plausible results without executing the actual work).
Beyond malicious behavior, integrity checks help to prevent faulty executions
(e.g., data corruption, hardware faults, or misconfigurations).
Ensuring computation integrity is essential, but is rarely considered in existing encrypted databases.
Computation integrity can be achieved by requiring the server to provide a proof that the encrypted result was computed using the targeted data and function.
Along this line, we introduce a verification protocol that allows the server to validate the output of in-range aggregations over ciphertexts with a succinct tag that can be verified in constant time at the client. To generate the proof, we use homomorphic Message Authentication Codes (\mac)~\cite{catalano2013practical}.
While \macs have been introduced as cryptographic building blocks in the literature, existing solutions do not achieve integrity while maintaining scalability.

%\remove{Ensuring verifiable malleability can be done by requiring the entity performing the homomorphic operation to provide}
%\red{The challenge here is how to ensure integrity and scalable encrypted data processing at the same time.}
%\blue{Any query result should include}  a proof that the ciphertext was computed using the \blue{correct} data and function. 
%\remove{Along this line, We introduce a verification protocol that allows the server to validate the output of \textit{in-range aggregation} over ciphertext with a succinct tag that can be verified in \textit{constant time} at the client. }
%\blue{Analogously to the decryption of query results, we want to ensure that the clients can verify the correctness of large in-range aggregations in constant time.}
%\remove{
%The tags preserve data confidentiality, meaning that the
%server does not learn any information about the input data nor the query result.}
%
%
%\fakeparagraph{Homomorphic Integrity.}
%\red{To achieve this, we augment the ciphertexts with \textit{homomorphic Message Authentication Codes (\macs)}~\cite{catalano2013practical}.}

%\blue{We augment the ciphertexts with  }\textit{homomorphic Message Authentication Codes (\macs)} ~\cite{catalano2013practical, hom-mac-boneh, first-hom-mac-mpc}.
\fakeparagraph{HoMAC.} 
Conventional Message Authentication Codes (MACs) are small tags generated for each ciphertext
which later ensure the authenticity and integrity of the ciphertext.
%They can then later be used by the client to authenticate the ciphertexts presented by a server.
\emph{\macs}~\cite{catalano2013practical,hom-mac-boneh} are conceptually similar to MACs,
but additionally allow the server to perform computations like aggregations over the ciphertexts, and to produce new tags that authenticate the outputs of the computation.
%\blue{This property of \macs allows the server to perform an aggregation over ciphertext/tag pairs ($(c_i, \sigma_i), ..., (c_j, \sigma_j)$) to generate a new tag $\sigma$ that authenticates the aggregated ciphertext $c$.}
%This property of the homomorphic MACs allows the server to generate proofs of correct evaluation on encrypted data without the need to decrypt or access \mac keys.
%\blue{Principals (e.g., data consumers) with access to the corresponding keys can \textit{verify}
%the validity of the ciphertext/tag pair $(c, \sigma)$ without the need to know the input data.
%A succeeding check guarantees that the output $c$ corresponds to the aggregation over the all the ciphertexts from $i$ to $j$.}
%We now give a brief overview on \macs and then introduce our protocol.
%Similar to standard message authentication codes (MACs), \macs authenticate data blocks but additionally support linear operations on the tag. 
%In \oursystem, for each data $m_i$, the client generates the ciphertext $c_i$ and \mac tag $\sigma_i$.
More precisely, the client generates a \mac tag $\sigma$ for each ciphertext $c$ and uploads $(c,\sigma)$,
where $\sigma$ is defined as follows:
\begin{equation}
\sigma = \mac_{s}(c) =  (s - c) / Z \mod p 
\end{equation}
where $s$ is a per-ciphertext key, $Z$ the \mac key, and $p$ a prime number.
%The server \blue{can simply}  %function $f$ \blue{over an interval $[i,j]$} on both encrypted data 
%\blue{over both ciphertexts} ($c_i,...,c_j$) and \mac tags ($\sigma_i,...,\sigma_j$).
The server computes aggregations on both the ciphertext and \textit{\mac} tags $\sum_{i=0}^{n-1} (c_i, \sigma_i) = (c_{res}, \sigma_{res})$.
The resulting tag $\sigma_{res}$ authenticates and verifies that the output $c_{res}$ corresponds to that specific aggregation.
A client in possession of the \textit{\mac} key material can verify the result by checking that the received $\sigma_{res}$  tag matches the ciphertext $c_{res}$:
\vspace{-10pt}
\begin{equation}
\sum_{i=0}^{n-1} s_i  \stackrel{?}{=} c_{res} + \sigma_{res} Z  \mod p
\end{equation}
% $f$ \blue{over the ciphertexts} from $i$ to $j$.
\macs are interesting for our use-case, since their symmetric nature makes them appealing to integrate with \enc. \newline
In contrast to authenticated data structures~\cite{Verena,integriDB-verification}, which can be used for outsourced computation verification, 
\mac tags do not need to be updated when new data is inserted.
However, without further optimization, their verification overhead prevents their use in our setting.

\fakeparagraph{Integrity Protocol.}
While \macs provide the desired integrity guarantees, they
%However, \macs
suffer from a verification overhead that is linear in the number of records in the aggregation query. % (e.g., checking a tag over 100 ciphertexts would require 100 key derivations and aggregations). 
Therefore, we apply a similar key canceling technique as already discussed above in the context of encryption:
%
%
%
%
%For each \blue{type of } digest in a stream,  ... a unique \textit{\mac} key $Z$ and ...
We define a  \textit{\mac} keystream $\{s_0, s_1, s_2, ...\}$ %, all of which are elements in $\mathbb{F}_p$. 
%Note that each key in the keystream and each message are labeled with a unique identifier $i$.
%
%The \mac suffers from the same problem as Castelluccia encryption (\rsec{crypto:primitives}).
%The client has to perform the same computation as the server.
%To overcome this challenge, we apply the key-canceling encoding.
%\blue{with the key} $s'_i = s_i - s_{i+1}$:
% $c_i = Enc_{k'_i} (m_i)$,
and, for each ciphertext $c_i$, the client computes the \textit{\mac} tag $\sigma_i$ as follows:
\vspace{-2pt}
\begin{equation}
 \mac_{s'_i}(c_i) =  (s'_i - c_i)/Z=(s_i - s_{i+1} - c_i)/Z\Mod p 
\end{equation}
Setting $s'_i = s_i - s_{i+1}$ %to compute the tag allows us to apply the  key canceling technique  discussed above.
enables a constant time verification at the client side regardless of the input size, since only the two outer keys are required:
\vspace{-8pt}
\begin{equation}
\sum_{i=0}^{n-1} s'_i  = s_0  - s_n \stackrel{?}{=} c_{res} + \sigma_{res} Z  \mod p
\end{equation}
%The in-range aggregated ciphertext and tag tuple $\sum_{i=0}^{n-1} (c_i, \sigma_i$) has the form:
%\begin{equation}
%(c_{res}, \sigma_{res}) = (Enc_{k_0 - k_n}(r) = r + k_i - k_j, \sum_{i=0}^{n-1}  \sigma_{i} = s_0 - s_n + r Z \mod p )
%(c_{res}, \sigma_{res}) = (Enc_{k_0 - k_n}(r),   s_0 - s_n + r Z \mod p )
%\end{equation}
Using the key canceling concept in both encryption and integrity is a key enabler for our efficient cryptographic access control (\rsec{keymanagement}).
Since verification of aggregation results does not require access to the individual messages that were aggregated,
our integrity protocol also integrates well with the resolution-based access control (\rsec{access-resolution}).

%This ensures that the in-range aggregation was executed over the correct query type \blue{(i.e., digest)} and the expected time range.
%\todo{Sell that it's really cool that this can use the same key canceling technique, and sell that this is essential to making the clever KDF tree and AC stuff work!}

\fakeparagraph{\mac Security.}
For an attacker, it is computationally infeasible to generate a forged ciphertext and a tag which pass the verification.
Note that we use different \mac key streams not just per-stream, but also per type of digest, i.e., target function.
Therefore, the server cannot substitute a digest aggregation with another.
In the case of key leakage,
a party with access to the \mac key $Z$ would be able to forge tags, but data confidentiality always remains intact.
For a complete security treatment of \mac we refer to~\cite{catalano2013practical, Aggregatable-prf} or the appendix~(\rsec{appendix}).

\subsecspacingtop
\subsection{Cryptographic Access Control}
\subsecspacingbot
\lsec{keymanagement}
%\todohs{The need to restrict access for different resolution levels was  already discussed several times.}
The symmetric homomorphic encryption and \mac both require a pseudorandom keystream with one key for each message.
The conventional approach to efficiently generating such keystreams would be to leverage a pseudorandom function with an initially exchanged secret key.
%Castelluccia et al.~\cite{castelluccia2009} propose leveraging a pseudorandom function~(PRF) with an initially exchanged secret key, to generate the keystream $\{k_0, k_1, k_2, ...\}$.
%Other systems~\cite{castelluccia2011CancelOut, seabed, bonawitz2017practical} adopt this approach and use a PRF $F(k, i)$ (e.g., a hash function or\linebreak
%a block cipher) to  generate the $i$-th encryption key based on a secret key $k$.
This allows handling a large number of keys with one secret.
%However, it exhibits the all-or-nothing sharing principle, as with access to the secret key $k$ one can compute all keys. 
However, with this approach, one could only share the entire data stream (i.e., all-or-none or in other words no fine-grained access control).
Instead, we want to allow efficient sharing of arbitrary intervals, and want to allow users to restrict access to lower-resolution data, e.g., hourly or daily summaries. 
To realize this granular access control and to allow data owners to cryptographically enforce the scope of access to their data, we design a novel key derivation construction.

\subsubsection{Key Derivation Trees}

\global\csname @topnum\endcsname 0
\begin{figure}[t]
	\center
	\includegraphics[width=1\columnwidth]{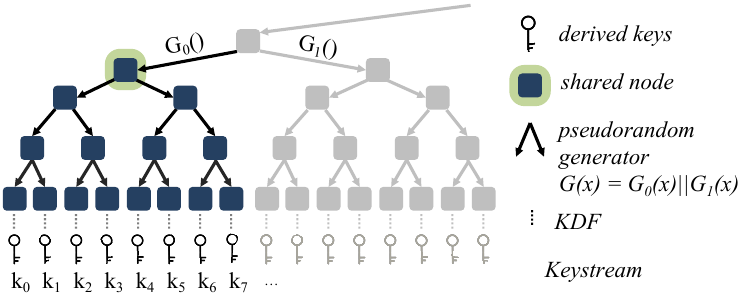}
	\vspace{-20pt}
	\caption{
		\setsmallcaption
		\oursystem's key derivation tree (leafs form a keystream).
	}
	\vspace{-15pt}
	\lfig{ht:keys}
\end{figure} 
Our key derivation is based on \emph{key derivation trees}, i.e. balanced binary trees where each node contains a unique pseudorandom string.
The leaf nodes represent the inputs to a key derivation function (KDF) to compute the keystream $\{k_0, k_1, k_2, ..., k_{2^h-1}\}$ as depicted in \rfig{ht:keys}.
The key derivation tree is built top-down from a secret random seed as the root.
The child nodes are generated with a pseudorandom generator (PRG) \shorten{- formally defined \red{in~\rsec{appendix:cryptodefinitions}} -}
that takes the parent string as the input.
Our PRG consists of $G_0(x)$ for the left-hand child and $G_1(x)$ for the right-hand child, where $x$ is the parent node. 
This procedure is applied recursively until the desired depth $h$ in the tree is reached. 
We select a large $h$ such that the keystream is virtually infinite, especially when considering that high-frequency streams will be chunked into e.g., one chunk per second.
%\todo{mention chunking}
%\todo{we can remove all this text.}
The pseudorandom generator can be realized from hash functions $G_0(x) = H(0||x), G_1(x)= H(1||x)$ 
\shorten{or block ciphers $G_0(x) = B_x(0), G_1(x) = B_x(1)$} with $x$ as the key.
%In~\rsec{eval:client}, we discuss the trade-offs of different PRGs and why AES-NI is the best candidate in terms of performance.} 
%For a formal treatment and security proofs of our key derivation tree, \blue{we refer to the extended paper~\cite{timecrypt_extended}}. %~\rsec{prooftree}.
%To compute a key from the root seed, at most $h$ function evaluations are needed. 

%
%For instance, in a binary hash tree, two hash functions (for left and right child nodes, respectively) are consecutively applied until the desired depth in the tree is reached, as depicted in \rfig{ht:keys}.
%The leaf nodes are fed to a key derivation function (KDF) to compute the per-chunk master encryption keys~(MEK).
%From the MEK, we derive the corresponding encryption keys for encrypting a chunk and its \chunksum.
%Note that we use two different KDFs to derive different keys from the same node for encryption of a chunk and its \chunksum, to avoid using the same key twice.
%For our encryption scheme, the KDF reduces the output to a 64-bit key with XORing 64-bit blocks.
%In \oursystem, we employ a $k$-ary tree of \blue{depth $d$} with $H(j||node)$ and $j \in [1,k]$, to realize $k$ different hash functions.
%In~\rsec{eval:client}, we discuss the trade-offs of different hash functions and why AES-NI is the best candidate, when used as a pseudorandom permutation.

\fakeparagraph{Access Token.}
The key derivation tree allows us to share segments of the keystream efficiently. 
Instead of sharing the segment key-by-key, the client shares a few inner tree nodes, combined into an \emph{access token}.
%A principal, who is in possession of these tokens, can derive the keys in the segment by computing the subtrees of each token, as illustrated in \rfig{ht:keys}.
%Therefore, for sharing a segment in the keystream, the client has to send at most $h$ access tokens instead of the individual keys, 
For instance in \rfig{ht:keys}'s toy example, a data owner grants access to the stream from $t_0$ to $t_7$,
and the corresponding key segment $\{k_0, \dots, k_7\}$ is shared using a single node. 
In practice, a single node in the tree can be used to share thousands of keys.
Note that given a node it is computationally not feasible (i.e., due to one-way property of PRGs) to compute the parent, sibling, or any of the ancestor nodes.
Hence, a data consumer cannot compute any keys outside the segment they are granted access to.
%Access tokens are encrypted per principal's public key (hybrid encryption) and stored at the server.

%\todo{These tokens can be shared with other users, we explain in section ...}

\fakeparagraph{Token Distribution.} 
Once the data owner specifies an access policy for a data consumer, the
\oursystem client generates
an access token which encapsulates the inner nodes of the tree needed to derive the corresponding shared keystream segment specified in the access policy.
We use the same key derivation tree for the encryption and \mac keystreams, but with a different KDF\footnote{Each leaf node of the primary key derivation tree is used to produce cryptographic keys needed for its corresponding chunk. Namely, keys for each element in the digest (i.e., query type), chunk, and \mac. Hence, we use different KDFs with the same node.}.
The token also contains encoded information about the subtree height and key identifier offset.
\oursystem then encrypts the tokens with the data consumer's public key (i.e., hybrid encryption) and stores it at the server, such that the data consumer can fetch it to gain access to the keying material required to decrypt the data or query results.
Note that \oursystem's key distribution is pluggable and we can employ alternative solutions. 
For instance, we can encrypt the token with attribute based encryption~\cite{Sieve} to share tokens based on attributes (e.g., month as a key attribute).
\subsubsection{Resolution-based Access Restriction}
\lsec{access-resolution}

%\vspace{-3pt}
%\todo{Feels redundant with content in Section 2 - Amy}
%\todo{Briefly summarize the goal here (to enable access policies that operate at different resolutions), move all of the explanation why that's a good idea to section 2?}
We now discuss how \oursystem provides  crypto-enforced access control over the \emph{resolution} at which data can be queried;
i.e., the data owner not only restricts access to a time range per data consumer but also defines the temporal granularity (e.g., per minute) at which they can retrieve or query data.

\fakeparagraph{Resolution Levels.}
\lsec{outer-key-sharing}
In \oursystem, the highest resolution for queries and access control is defined by the chunk size $\Delta$.
%	\footnote{However, the underlying raw data might be even higher-resolution.}.
Whenever we aggregate over an interval, we reduce the data resolution.
For example, with one second chunks, an aggregation over 60 chunks results in a per-minute resolution.
We can exploit the fact that keys cancel out during in-range aggregations, as described in~\rsec{efficient-dec}, to cryptographically restrict access to lower resolution levels.
In general, a ciphertext generated through an in-range aggregation over the time period $[t_i, t_j)$ has the form: 
\begin{equation}
\sum_{x=i}^{j-1} c_{x} = \sum_{x=i}^{j-1}  m_{x}+ k_i - k_j
\end{equation} 
where the inner keys are canceled out.
Hence, given access to just the boundary keys $k_i$ and $k_j$, one can decrypt the aggregation, % $\sum_{x=i}^{j-1}  m_{x}$
but none of the individual ciphertexts.
Resolution levels must be multiples of the chunk size $\Delta$ and the segments at a given level must not overlap.
%Resolutions \blue{must be} aligned at timestamps and cannot be shifted arbitrarily.
%For instance, per-minute resolution means only aggregated data at the full minute from the given timestamp.
Otherwise, data consumers could compute the difference of two aggregates overlapping by e.g., one chunk, allowing them to learn the data for that chunk which would violate the resolution-based access policy.
For example, if the data owner wants to restrict access to a 3-fold resolution of the chunk size, the data owner would share only $\{k_0,k_3,k_{6}, ...\}$ with the data consumer.
The data consumer can then decrypt the aggregated ciphertexts at the 3-fold (i.e., $3\cdot\Delta$) or lower resolutions, but cannot access higher resolutions since the inner keys are missing.
\begin{figure}[t]
	\center
	\includegraphics[width=1\columnwidth]{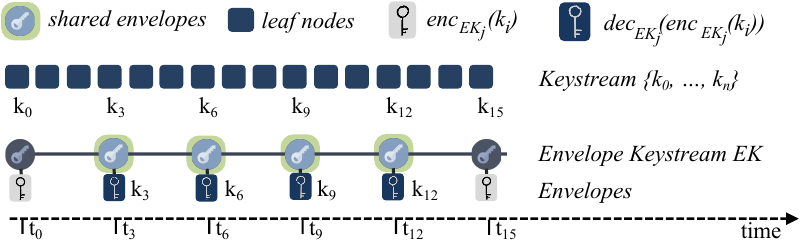}
	\vspace{-15pt}
	\caption{
		\setsmallcaption
		Envelope encryption for resolution-based access, showing envelopes required to share $[t_3,t_{12}]$ at a resolution of $3\Delta$.}
		%Each resolution level has a separate envelope keystream.}
	\vspace{-15pt}
	\lfig{leveled:access}
\end{figure} 

\fakeparagraph{Envelopes.}
While a data owner could share the boundary keys required for resolution-based access directly, this is not efficient since the number of keys necessary is linear in the length of the shared interval.
Instead, the data producer stores the required boundary keys for a stream on the server, protected by another layer of encryption, the \emph{envelope}.
The keys used for the envelope encryption are derived from a new tree-based keystream $\{\bar{k}_0, \bar{k}_1, \bar{k}_2, \ldots\}$.
For each resolution level, we use a different keystream for the envelope encryption.
For example, if a data owner wants to make a per-minute resolution available for a stream with 20~s data chunks, the data owner encrypts the boundary keys $\{k_0, k_{3},k_{6},\ldots\}$ with the envelope keystream, and stores $\{enc_{\bar{k}_1}(k_0), enc_{\bar{k}_{2}}(k_{3}), enc_{\bar{k}_{3}}(k_{9}),\ldots\}$ on the server, as shown in~\rfig{leveled:access}.
%
%
%
%The resolution-based access control requires sharing keys at a fixed interval, which grows linearly in size.
%Resolution keys consist of a set of keys $S_r$ sampled at a fixed time interval --- defined by the resolution level -- from the data encryption keystream.
%To be able to share a temporal segment of resolution keys $S_r$ efficiently, we generate a tree-based keystream $\{\bar{k}_0, \bar{k}_1, \bar{k}_2, ...\}$ for each resolution level. 
%We encrypt keys in $S_r$ with the corresponding resolution keystream.  
%
%\oursystem encrypts the interval keys $\{k_6, k_{12}, .., k_{42}\}$ with the keystream $\{enc_{\bar{k}_1}(k_6), enc_{\bar{k}_{2}}(k_{12}), .., enc_{\bar{k}_{7}}(k_{42})\}$ 
%and stores the resulting envelopes, indexed by their timestamp and resolution, on the server, \blue{as shown in \rfig{leveled:access}}.

Sharing a stream at a lower resolution is then again a matter of sharing a single access token, with the difference that
the token now contains nodes of the key derivation tree for the envelope keystream, rather than for the original encryption keystream.
A lower-resolution query returns, in addition to the encrypted result, two envelopes containing the two boundary keys required to decrypt the aggregated ciphertext.
% which the client decrypts using the shared access token.
The overhead of resolution-based access control is similar to access control without resolution restrictions (\rsec{keymanagement}), i.e., an access token consist of at most $O(log(n))$ nodes from the key derivation tree.
\begin{table*}[t]\centering
    \scriptsize
    \begin{tabularx}{\textwidth}{>{\hspace{0.4pc}} l >{\hspace{0.8pc}}  l }
    \toprule

    \textbf{Function} & \textbf{Description} \\
    \midrule
%    \midrule 
%\hline
              (1)~\texttt{CreateStream(uuid, [config])} &  Create a new stream,
                                         config defines parameters, e.g., chunk interval, operators.\\         
         (2)~\texttt{DeleteStream(uuid)} &  Delete specified stream with all associated data.\\      
         (3)~\texttt{RollupStream(uuid, res, [$T_s$, $T_e$])}  & Rollup an existing stream or a segment of it to the specified resolution.\\
         \midrule 
%\hline
              (4)~\texttt{InsertRecord(uuid, [t, val])}  & Serialize data points in a chunk and append to the end of the stream.\\ 
         (5)~\texttt{GetRange(uuid, $T_s$, $T_e$)}  & Retrieve all data records within the specified time interval.\\   
         (6)~\texttt{GetStatRange([uuid], $T_s$, $T_e$, resolution, [operators])}  & Retrieve statistics for the given time interval and resolution,
default [sum, count, mean, var, freq]. 
\\    
         (7)~\texttt{DeleteRange(uuid, start, end)} & Delete specified segment of the stream, while maintaining per-chunk \chunksum. \\     
         \midrule
%\hline
              (8)~\texttt{GrantViewAccess(viewid, [princ-id])} &Grant access to an existing View.  \\
              (9)~\texttt{CreateView(viewid, [policy])} & Create a View with the given policy in JSON format.\\
              (10)~\texttt{CheckView(viewid, princ-id)} & Retrieve a View token.\\
   %      (9)~\texttt{\red{GrantOpenAccess}(uuid, principal-id, start, res)}   & Grant open-ended subscription, i.e., access granted until revoked. \\    
         %(9)~\texttt{Revoke\-Access(uuid, principal-id, end)} &  Revoke access of a principal starting from the specified end time. \\    
    \bottomrule
%\hline
    \end{tabularx}
    \vspace{-8pt}
    \caption{\setsmallcaption\oursystem's basic API.} 
     \ltab{API}
    \vspace{-15pt}
\end{table*}
%%%%%%%%%%%%%%%%%%%%%%%%%%%%%%%%%%%%%%%

\fakeparagraph{Dynamic Resolution Levels.}
In \oursystem, 
a user does not need to decide a priori on a fixed resolution for data consumers and can dynamically at any point in time define a new resolution.
E.g., Alice can share her health data with a physician at minute-level (high-resolution) during physiotherapy from Jan-to-Feb, and from March reduce the resolution to hourly (low-resolution).
The physician only sees high-resolution data for Jan-Feb and only hourly-data from March onwards.
\subsecspacingtop
\subsection{Access Control Extensibility}
\subsecspacingbot
\lsec{extensibility} 
  %\vspace{-5pt}
Beyond temporal and resolution-based access policies, our construction also lends itself to enabling privacy policies on encrypted data, as combining ciphertexts from multiple users creates valid ciphertexts under a new \textit{virtual aggregate key}.
In the context of private operations, privacy policies permit a data consumer (e.g., analyst) to only run cross-stream aggregate queries, without having access to individual data streams.
Similar to data access policies, privacy policies in our system are enforceable via encryption. 
As a concrete example, a user might want to allow a research lab to query her data but only if aggregated with a fixed set of $n$ users, to preserve her individual privacy.
%and that such aggregates shall leak no information about whether her data is part of the aggregate or not (differently-private aggregation).
%For brevity, we only discuss the first case, in principle, the latter can be realized in a similar manner, however, in combination with other techniques~\red{\cite{castelluccia2011CancelOut}}.
Ensuring that a data consumer can only decrypt aggregates across a set of users can be realized by ensuring that she only has access to the \textit{virtual aggregate key}
(i.e., the data consumer never sees the keys for a particular user's stream in isolation). 
For instance, if a service is authorized to access an aggregate query over $n$ encrypted messages from different users,
then sharing only the \textit{virtual aggregate key}  $\sum_{i=1}^{i=n} k_i$ will ensure that the analyst can only decrypt the aggregated result.
Therefore, we need a way to compute the \textit{virtual aggregate key} without exposing the individual keys $k_i$ of each user to any of the involved parties; the storage provider, authorized data consumer, or other users. \newline
This can be accomplished by a secure aggregation protocol~\cite{bonawitz2017practical,castelluccia2011CancelOut} between the involved users and the analyst.
The inputs to the protocol are the users' individual keys $k_i$ and the output is the blinded contributions towards the \textit{virtual aggregate key}.
Queries across streams can be performed efficiently on the server, and the analyst can only decrypt the final result  
via a \textit{virtual aggregate key}.
In~\rsec{evaluation}, we discuss a private crowdsourcing application atop of \oursystem that uses this technique.
\secspacingtop
\section{Fast Analytics and Processing}
\secspacingbot
\lsec{index} 
\lsec{functions}
	%\vspace{-5pt}

%In the following, we discuss several system design aspects of \oursystem.

 \begin{figure}[t]
	\center
	\includegraphics[width=0.98\columnwidth]{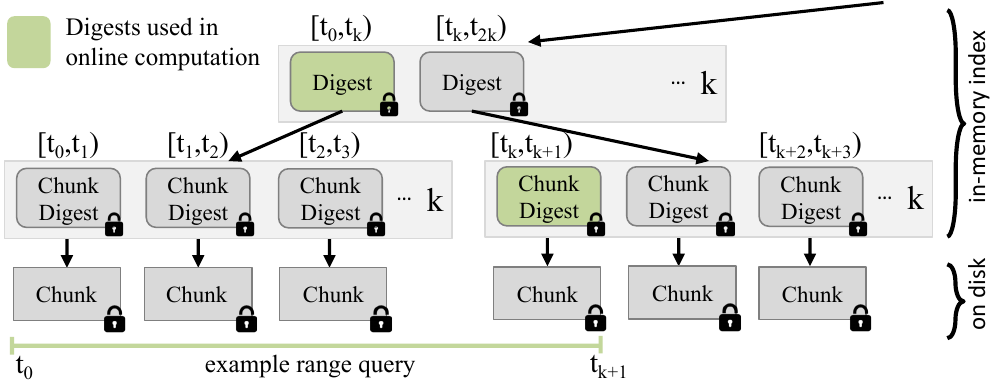}
	\vspace{-5pt}
	\caption{
		\setsmallcaption
		A statistical index for time series data with a $k$-ary time-partitioned aggregation tree.
		The pre-computed encrypted index allows for fast response times for statistical queries.
		%    whereas the large high-resolution time series data is stored on disk.
	}
	\vspace{-15pt}
	\lfig{time-partitioned-tree}
\end{figure}

%\todo{start this section by reiterating the main challenges of making queries/processing fast}
To meet the requirements of time series databases, \oursystem must handle massive amounts of data, yet at the same time be able to serve queries with low latency.
We address this challenge by introducing efficient client-side serialization/encryption and efficient encrypted indices on the server.

%In this section, we describe functionality supported by \oursystem over encrypted data.
%\oursystem supports common queries supported in existing time series 
%databases~\cite{chronix,BTrDB,summarystore,pisa,influxdata}, 
%such as statistical, select, and predict queries. 

\fakeparagraph{Client-side Data Serialization.}
%Why we did not take the effort to fully integrate \oursystem in an existing TSDB?}
%To meet the performance and end-to-end encryption requirements, the client takes an active role in \oursystem.
The client serializes and encrypts data chunks containing the raw data, and \chunksums.
The content of a \chunksum \shorten{$m = \{x_0, ..., x_n\}$} is set per stream based on the 
supported queries.
The \emph{default} query configuration of \oursystem supports $sum$, $count$, and $mean$.
Other query types such as $variance$, $standard$ $deviation$, $histogram$, bucket $min/max$, approximated $quantiles$,  $trend$ $detection$, and limited $filter$ queries, can be enabled.

%The \chunksum is encrypted with \enc. %- additively homomorphic.
%which provides semantic security, such that no information about the plaintext data is leaked.
%\red{The data points per chunk are compressed\footnote{\oursystem runs the compression algorithm that yields the best results for the underlying data.
%For instance, delta encoding might be highly effective for low precision data with many identical values, but less effective for high precision data with large deltas.
%Since the performance of compression algorithms varies based on the data stream type and characteristics, 
%\oursystem supports various lossless compression techniques, with \textit{zlib} as default.}
%}
%and encrypted with a randomized encryption scheme (AES-GCM-128).
%Next, we describe our encryption scheme, and how encryption keys are generated and managed for both chunk digest and chunk data encryption to enable flexible secure access control.

\fakeparagraph{Server-side In-memory Encrypted Index.}
\oursystem's server maintains an in-memory encrypted index based on a time-partitioned aggregation tree over encrypted data. 
This is a key building-block that enables us to serve low-latency analytics on large encrypted data streams and enables efficient data retention.
%To keep computation tractable over massive amounts of data,  
%\oursystem 
%This forms the base for an encrypted index that makes it possible to execute fast queries over a large sequence of encrypted data, 
%\red{The encrypted data computation is possible due to our homomorphic encryption construction HEAC.}
%\todoah{talk about why these indexing is important. prune search space not need to serial scan of the data.}
The index structure is a $k$-ary tree, where each internal node (digest) holds $k$ statistical summaries of the subtree below it. 
%\blue{Smaller values of $k$ speed up computation at the cost of additional memory usage.}
%\blue{A digest includes a statistical summary over a fixed time interval. The index is built bottom-up, from the encrypted chunk digests uploaded by clients.The chunk digests form the leave nodes, and an index inner node holds the digest which aggregates the leave nodes}
% and references a child node with $k$ more fine-grained summaries of the same time interval.
%Hence, the server can compute a digest in a tree node by recursively aggregating the digests in the referenced child node.
The tree leaves store the chunk digests encrypted with HEAC at the client and represent the highest resolution data summaries 
(\rfig{time-partitioned-tree}).
%The server builds the tree bottom-up:
On the arrival of a new chunk digest, the server inserts it as a leaf node, and updates statistical summaries of the parent 
nodes's by performing an encrypted aggregation.
Any operation that can be expressed as an aggregation of the intermediate results from the child subtrees can be included in the 
summaries (see \rsec{functions}).
%We discuss later the set of operations we currently support in \oursystem. 
Time series workloads are in-order and append-only, therefore updating the tree is straightforward. % for the server.
%the parent node represents the statistical summary spanning over the whole time interval of its subtree, as depicted in \rfig{time-partitioned-tree}.
%A digest is a summary of their corresponding sub-tree (i.e., the parent nodes contain the result of the aggregation of all digests in their child node).
%As illustrated in \rfig{time-partitioned-tree}, the chunk \chunksum of each time interval is stored in the corresponding leaf node, while the inner tree nodes hold an aggregated \chunksum over their subtree.
The encrypted index enables \oursystem to significantly decrease the response time for statistical queries, as the server avoids expensive serial scans. %and the pre-computed statistical summaries in the tree .
When executing a statistical range query over a time interval, % $[t_0,t_{k+1})$ (\rfig{time-partitioned-tree})
the server traverses the tree and selects only %to the depth and offset corresponding to the
the digests required to cover this interval, as illustrated in \rfig{time-partitioned-tree}.

\fakeparagraph{Statistical Queries.}%\todohs{this section lower in quality and would benefit from a rewrite}
%\todo{We need to revise to generalize this section from limited it to statistical queries.Maybe rather that we support linear operation. This inherently means that we support many learn statistical queries. But any other computation that are linear (e.g., linear ML models) and maybe point that there exist body of work that focus on transformation of such computation to linear to fit within MPC paradigm.}
%search (temporal range), sum, mean, variance, standard deviation, frequency count, and min/max.
%\red{These are , which are also essential for identifying data segments that are relevant for further 
%complex analytics.}
So far, we have developed the means to evaluate aggregates over ciphertexts, now we briefly\footnote{Due to space constraints, we keep the description here brief and refer to~\cite{prio,SEC,Rencher} for detailed description.} discuss how we combine aggregation with known encoding techniques~\cite{prio,SEC,Rencher} to allow \oursystem to compute more sophisticated statistics over ciphertexts.
%We now discuss the \chunksum encodings that enable \oursystem to process these queries over encrypted data.
At a high level, %in our setting, 
each per-chunk \chunksum holds a vector of encoded values % $\{x_0, ..., x_n\}$
that are encrypted with \enc.
%The server computes the aggregate function on the encrypted encodings across different digests to serve statistical queries.
For example, this vector might include the encrypted  \textit{sum} and \textit{count} of the data points in the chunk.
From this, we can then also calculate the \textit{mean}. 
%The resulting two values allow us to additionally compute the \textit{mean}.
To compute quadratic functions, e.g., \textit{var} and \textit{stdev}, the vector includes the \textit{sum of squares} of the points in the chunk.
% \blue{and frequency count} .
We can also include the \textit{frequency count} of data points in the chunk, 
which yields valuable information to compute several statistical functions, such as \textit{min, max, top N, bottom N, histograms, and quantiles.}
For frequency counts, we % either directly %here data values range over $[0,B]$ we 
use a vector  $[c_{v_1},.., c_{v_n}]$, %of length-$B$ the set of possible data values %$D = \{0, .., B-1\}$ 
where each element in the vector $c_{v_i}$ tracks the count of data points with value $v_i$.
This works well for small $n$, which is often the case for (discrete) time series data.
For larger ranges of values, we approximate the frequency count, i.e., each $c_{v_i}$ tracks the count of a small range (\emph{bin}) around $v_i$~\cite{prio}.

\fakeparagraph{Advanced Analytics.}
In principle, %any operation that uses intermediate aggregates from the subtrees of the encrypted index.
any operations with aggregatable transformations %meets this requirement,
can be supported in \oursystem, 
including a variety of sketch algorithms~\cite{sketch}.   
%The set of supported statistical queries in \oursystem can be extended with further aggregation-based functions.
In addition, we can support many forms of machine learning, %blue{with} aggregation-based transformations 
%can be supported ,
e.g., via aggregation-based encodings that allow private training of linear models~\cite{prio,SEC,Rencher}.
These types of analytics are often employed in time series data to understand and detect runtime anomalies, trends, and patterns. 
%Hence, they allow users to quickly identify and diagnose problems as they arise.
%forecasting
We show how such analytics can be realized in \oursystem, using the example of \textit{private trend detection},
i.e., identifying a general tendency over a defined time interval. %\blue{that} can be leveraged in anomaly detection.
It allows users to estimate the magnitude of a trend %(i.e., an indication for important changes) 
and is a highly related task to event detection (e.g., runtime anomalies).
Linear regression using least-squares is a simple yet powerful method for trend detection~\cite{ts-analysis}.
%\footnote{Assuming that the residuals do not correlate}. 
To compute a linear regression model over a stream, %$S$ observations %in the stream 
%[$(t_0, x_0)$, \ldots, $(t_S, x_%S)$] with $n$ records per chunk,
the per chunk digest is defined as ($\sum_i x_{i}$, $ \sum_i t_{i} x_{i} $) 
for $i \in [0,n)$.
This way the expensive aggregations are done at the server. % and the client computes the model parameters locally.}
Such learning on summarized data also delivers privacy gains, as the raw data is not exposed in the training phase.
%without impacting the accuracy of the model.}
In \rsec{evaluation}, we discuss the performance aspects of implementing such applications atop \oursystem.
\fakeparagraph{Filter Queries.}
\oursystem supports filter queries with predefined predicates.
One can define digest encodings that contain statistics over the values of the underlying chunk filtered with a predicate $P$ (e.g., the sum of all values larger than 10).
%\red{To retrieve an aggregation result over a time range of values filtered by $P$, the client fetches the aggregation of the respective encodings in the digest.}\todohs{it is still difficult to understand how it works. Not that you should give a detail intro to this concept, but at least rework the part about client processing part. If we fail to convey our message, then we should remove the paragraph and avoid confusing the reader.}
In the query phase, the filtered digests are used to compute statistics over the values matching the predicate $P$.
% (e.g., calculate the sum of all values that were above 10 last week).
%\red{The filter queries incur additional client-side preprocessing and storage overhead at the server.}}
%\todo{I would remove the last sentence. Support for any additional query in our system will incur such overhead. So does not make sense to mention this only here.}
\fakeparagraph{Time-Decayed Data Processing.}
As time series data ages, it is often aggregated into lower resolutions for long-term retention of historical data, 
while high-resolution data is aged-out.
Typical strategies %for collapsing multiple data points for roll-up to a lower precision 
are based on compact summaries through aggregates~\cite{summarystore, timescaleRetention, influxRetention}.
%Depending on the application, after a defined retention period.
\oursystem natively supports these approaches: %archiving at lower resolutions for encrypted data.
as our index maintains aggregated summaries of the raw data, 
%\oursystem 
we can selectively delete aged-out raw data and prune lower nodes in the index. 
For example, we implement a retention policy based on the time-decayed merge algorithm~\cite{summarystore} % as follows
which keeps the data store compact (logarithmic in the input size) 
by dynamically re-compacting older data as new data arrives.

\begin{table*}[t]
  \centering
   \footnotesize
  \begin{tabular}{lcccccccc}
%    \toprule
\hline
        \multirow{2}{*}{System}    &    \multicolumn{1}{c}{Micro}             &  \multicolumn{1}{c}{Index - Size}    & \multicolumn{3}{c}{Average Ingest Time}     & \multicolumn{3}{c}{Average Query Time (worst-case)}\\    
    \cmidrule(lr){2-2}\cmidrule(lr){3-3}\cmidrule(lr){4-6}\cmidrule(l){7-9}
                    & ADD          & $1M$         & $1k$    &$1M$ &$100M$        &$1k$ &$1M$   &$100M$\\ 
    \midrule
 %   Paillier~(used in CryptDB~\cite{cryptdb})                     & 2.1$\mu$s     & 780MB (96x)     & 37ms         (6167x)&     42ms  (3500x)& N/A                    & 22ms         (1692x)&         37ms (1028x)&  N/A\\  %36.6ms 21.9ms
   % EC-ElGamal~(used in Talos~\cite{talos})       & 0.7ms         & 168MB  (21x)        & 27ms        (4500x)&  43ms     (3583x)& N/A                  & 66ms        (5077x)&  185ms    (5139x)&  N/A\\    %26.8ms     %42.5ms % 65.6ms %184.6ms
      \textbf{\oursystem}                 & 1ns        & 8.1MB (1x)             & 10$\mu$s    (1.7x)& 16$\mu$s     (1.3x)& 22$\mu$s (1.3x)    & 21$\mu$s    (1.6x)& 46$\mu$s (1.3x)& 50$\mu$s    (1.1x)\\        %15.6us 21.5us 34.6
      \textbf{\oursystemPlus}                 & 3ns        & 24.3MB (3x)             & 16$\mu$s    (2.6x)& 35$\mu$s     (2.9x)& 39$\mu$s (2.3x)    & 38$\mu$s    (2.9x)& 87$\mu$s (2.4x)& 109$\mu$s    (2.4x)\\        %15.6us 21.5us 34.6

    Plaintext        & 1ns        & 8.1MB (1x)                 & 6$\mu$s     (1x)& 12$\mu$s     (1x)& 17$\mu$s     (1x)    & 13$\mu$s    (1x)& 36$\mu$s     (1x)&  45$\mu$s    (1x)\\ %12.7us  33.2us 55.3
%    \bottomrule
\hline
  \end{tabular}
   \vspace{-5pt}
% Memory discussion: (64*1000000-1)/(63) = 1015873 total nodes with 1M leaves
% 64bit * 1015873 --> 8.13MB
% 5x CRT * 33 Byte *  1015873= 167.62 MB
% 3072 bit *2 * 1015873= 780.2
  \caption{
    \setsmallcaption
        Overview of evaluation results on the cloud, with 128-bit security, except for plaintext.    
%    The overhead ratio compares to the plaintext setting.
%    The absolute values depict average cases for three different index sizes.
    The largest index size with 100M chunks, represents  50~billion data points in our health app. 
    %is omitted for Paillier and EC-ElGamal due to excessive overheads.
    %\todo{Here we should have seabed and remove Talos.}
    }
   \vspace{-6pt}
  \ltab{complete-eval} 
\end{table*}

% !TEX root = ../paper_secure_TS.tex

%Alternatively, access can be of the open-ended subscription type, such that it is granted until revoked.
%To resolve the principal-id, we assume internal or external identity providers, such as KeyBase~\cite{keybase}. %or Blockstack~\cite{blockstack}.  services to be in place
\begin{comment}To protect against unauthorized inserts or query requests, we rely on authorization frameworks, such as OAuth~\cite{oauth}.
\end{comment}

\begin{comment}
\begin{figure}[t]
	\center
	\includegraphics[width=1.0\columnwidth]{images/indexVersioning}
		%\vspace{-20pt}
	\caption{
		\small
		A simple consistency technique for the index to ensure the latest version of an index node is in the cache. 
%		Each node holds a version number $v$, which represents how many elements are currently stored in the leaf nodes of the corresponding subtree.
%		In case the version of the node is outdated, \oursystem fetches the latest version of the node from the storage.
	}
%			\vspace{-10pt}
	\lfig{index:versioning}
\end{figure} 
\end{comment}

\shorten{
\fakeparagraph{Read, Write, and Query Stream Data.}
\blue{
  Data writers periodically append to streams by invoking \textit{InsertRecord}. 
  The \oursystem client hides the cryptographic operations such as key storage and encryption and adds the digests based on the configuration received upon stream creation. 
  Consumers can query the data streams with \textit{GetStatRange} providing a time range, the statistics to retrieve, and the resolution of aggregation in this time window.
Given the resolution access flag, the server additionally returns the necessary envelopes for. 
}}
\secspacingtop
\section{Prototype} 
\secspacingbot
%\vspace{-5pt}
\lsec{implementation}

%We built a prototype of the proposed system to evaluate its practicality. 
%In this section, we describe several interesting parts of our \oursystem prototype.

%\subsection{\oursystem Integration} 
%\lsec{design:datamangement}
%Here we need to describe what we opt to be our storage and data management systems.
%If this is not already outlined in the background justify our decision of opting for this model, and emphasize that this is merely a 
%realization. We think that \oursystem can be adapted to work with other systems. 
%Describe how these peaces are put together with examples (e.g., detail the API and give example queries).

%The core of \oursystem is an efficient encrypted time series indexing engine.
%This index is designed to boost compression, throughput, and query speed for raw time series data with strong level of confidentiality.  
%In this section we briefly describe storage, consistency, and integration aspects of our system and
%how they interplay with the encrypted index to realize an efficient encrypted time series database.

%The core of \oursystem is an encrypted index that facilitates executing efficient statistical queries  on time series data without disclosing plaintext data to the server.

%We briefly describe \oursystem's API and storage and integration aspects of \oursystem.
%We now describe the API exposed by \oursystem to application developer. 
\fakeparagraph{API.} \oursystem is realized as a service which exposes an interface similar to conventional time series stores~\cite{chronix,BTrDB,summarystore}; 
applications can insert encrypted data, retrieve encrypted data by specifying an arbitrary time range, and process statistical queries over arbitrary ranges of encrypted data, as summarized in 
\rtab{API}.
In \oursystem, each stream is identified by a unique UUID and
%associated with a collection of attributes (\textit{stream metadata}) which specify the type and source of data (e.g., hostname, data type, sensor ID, location). 
associated \textit{stream metadata}, e.g., hostname, data type, sensor ID, location. 
%We refer to this set of attributes as.
%More importantly, each stream comes with an associated set of \textit{per-chunk \chunksum} that determines the type of statistical queries that  can be carried out on the stream.
%The nature of chunk \chunksum can vary with respect to the intended application and data types.
%Queries are performed on temporal segments of the data stream.
Each stream has one writer (i.e., data producer) and one or multiple readers (i.e., consumers).
A data owner can grant and specify access polices to consumers. 

\fakeparagraph{Granting Access to Stream Views.}
Data owners can manage access to their stream resources with the View API. 
\textit{View}s define what a data consumer can access within the scope of the \textit{View}.
\textit{View}s are set in JSON format, containing a unique identifier and a list of per stream access policies. 
In the current version, an owner can define the time range and the granularity that is accessible per stream, as for example:
\textit{
"viewid": 2999,
"streams": [\{"uuid": [9,10], "from": "1546315200", 
"to": "1546315800", "granularity": "60s" \}]}.
This \textit{View} defines an access scoped to stream 9 and 10 in the specified time window with a minute granularity. 
After the user defines a policy, the API assembles the access token with the necessary inner nodes of the key construction for the specified \textit{View} (\rsec{keymanagement}). 
The client library then derives a \textit{View} key, encrypts the token along with the JSON description using AES-GCM and uploads it to the server. 
To give data consumers access to the \textit{View}, the client invokes the \textit{GrantViewAccess} command, which encrypts the \textit{View} key with the respective consumer's public keys.
The authorized consumers can download the tokens for the given \textit{View} and can query the streams in the defined scope by the access policy.
Though access policies are enforced by encryption, the intricacies of the key management are insulated from users in our design.

\fakeparagraph{Reference Implementation.} \oursystem's prototype is implemented in Java and consists of 6k SLOC 
with additional 4k SLOC for the applications and benchmark code.
%The client and server consist of 850 and 5.6k sloc, respectively.
We used \textit{Netty}~\cite{netty} for network communication. 
\oursystem's server and client communicate over Google's \textit{protobuffers}~\cite{protobuf} protocol.
The current prototype uses Cassandra~\cite{cassandra} as the storage backend.
%Our prototype resembles a common architecture of existing time-series databases~\red{\cite{}}, 
%where a time-series daemon is interfaced with a distributed key-value storage. ==> HS: we have already mentioned this before
%However, \oursystem's design is not coupled with this specific architecture, and to a large extent should work with any time-series database. 
%\blue{One could as well use \oursystem daemon in front of another on-disk TSDB.}
%should work on top of any distributed key-value storage.
The encrypted index is augmented with the in-memory cache~\textit{caffeine}~\cite{coffein} to speed up index node access.
%We employ an LRU cache for the index nodes using \textit{caffeine}~\cite{coffein}.
%The client library consists of data serialization module, which serialize data streams in into chucks.
%Client encrypts and sign data their data with AES-GCM and encrypt data digests with HEAC.
For the implementation of the cryptographic schemes, we used the Java security provider and a native C implementation of AES-NI.
%\red{We implemented HEAC using \red{XX}.}
%The query module implements the statistical query interface and a standard data retrieval interface.
%A user can query and access data only if she is granted the corresponding access permissions.
We compare the encrypted index performance with HEAC against alternative private aggregation schemes. 
We implemented three variants of the encrypted index based on Paillier~\cite{javalier} (Java 
\textit{BigIntegers}), EC-ElGamal (OpenSSL~\cite{openssl}), and ASHE (we implemented it as described 
in~\cite{seabed}).
%\footnote{We reached out to the paper authors to use their code in our evaluation, they informed us that the
%code is not ready to be released.}).
Our code is available online.

 %~\cite{ecelgamallib}
%For the Paillier implementation we rely on the \textit{javallier} library~\cite{javalier}, which is based on the Java \textit{BigInteger} primitive. 
%The EC-ElGamal implementation uses the elliptic curve operations of OpenSSL~\cite{openssl}.

% !TEX root = ../paper_secure_TS.tex
%\vspace{-5pt}
\secspacingtop
\section{Evaluation}
\secspacingbot
\lsec{evaluation}
  %\vspace{-5pt}
%\todo{be consistent with respect to Plaintext vs data in clear (plots, table, and text). We should make 
%the setup way more concise}
%\todo{fix ingest vs insert make consist in plots and text.}
%\todo{
%Lets formulate the eval questions:
%\begin{itemize}
%\item What are the induced overheads of encryption and access control?
%\item Does \oursystem scale to large scale workloads despite its security features? 
%\item Can \oursystem run compelling real-world applications?
%\end{itemize}
%}

%We have developed \oursystem to evaluate its practicality.
In this section, we evaluate \oursystem's practicality.
Our evaluation answers three core questions:
\textit{(1)}~Can \oursystem meet the performance requirements of time series applications?
%- by showing to which extent we narrow the performance gap of \oursystem compared to operating on plaintext data.
\textit{(2)}~What are the performance gains of HEAC compared to alternatives? --- 
HEAC supports access control and secure computation simultaneously;
both aspects have traditionally been addressed with different schemes,
consequently we examine alternatives independently in our evaluation.
\textit{(3)}~Can \oursystem run compelling real-world applications?    

\fakeparagraph{Setup.} Our experiments are conducted in Amazon AWS, on M5 instances
 equipped with a 2.5 GHz CPU\shorten{Intel Xeon processors}
running Ubuntu (16.04 LTS).
 \oursystem's server runs on an m5.2xlarge instance with 8 virtual processor cores (vcores) and 32~GB of RAM 
and a Cassandra node runs on an m5.xlarge instance with 4 vcores and 16~GB of RAM.
The clients are simulated on several m5.xlarge instances.
The client and server are located in the same data center network, with up to 10~Gbps bandwidth.
In the microbenchmark, we quantify the overheads of encryption and decryption on end devices. 
We consider resource-constrained IoT devices; this class of devices is a major source of sensitive time series data.
We use IoT OpenMotes (32-bit ARM M3 SoC 32~MHz) %and a 250~MHz  crypto accelerator
%(Fitbit utilizes a similar class of \mbox{micro-controllers})
%(specs: 32-bit ARM M3 SoC at 32~MHz, 32~KB  RAM, and a crypto accelerator running up to 250~MHz), 
%a mobile phone (Nexus 5, 2.3~GHz quad-core 64-bit CPU, 2~GB RAM) and 
and a MacBook Pro 2.8 GHz Intel Core i7, with 16~GB RAM.

%Unless noted otherwise, we use the same system architecture in our .
We quantify the overhead of \textbf{\oursystem} (confidentiality) and \textbf{\oursystemPlus} (confidentiality plus query verification), and compare it to 
\textit{(i)}~operating on \textbf{plaintext} as the baseline, %operating over data in the clear, i.e., 64-bit unencrypted values,
%\textbf{\oursystem} where we employ our encryption scheme to encrypt chunk \chunksum and use AES-GCM encryption for 
%encrypted the compressed chunks.
%\red{For the key-regression, we use the CPU-based AES-NI instructions.}
and
\textit{(ii)}~\textbf{prior work} where we consider alternative encryption schemes for encrypting the \chunksum,  
i.e., \textit{Paillier} (used in CryptDB~\cite{cryptdb}), \textit{EC-ElGamal} (used in Pilatus~\cite{pilatus}) and \textit{ASHE} (used in Seabed~\cite{seabed}).
For access control, we compare to a strawman solution and a construction of KP-ABE (used in Sieve~\cite{Sieve}), that we use to realize temporal access control similar to that supported by HEAC.    
%In the E2E evaluation, we compare \oursystem to Seabed~\cite{seabed} which uses Counter-based Castelluccia~\cite{castelluccia2009}.
\shorten{For fairness, in all settings, the system operates with compressed data chunks.}
Unless noted otherwise, we use 128-bit security~\cite{nistrecommendation}, i.e., 3072-bit keys for Paillier and 256-bit elliptic curves for EC-ElGamal (i.e., prime256v1). % in OpenSSL).
For the microbenchmarks, we use synthetic large data that resembles the mhealth application (\rsec{eval:e2e}) dataset.  

%In the microbenchmarks, we compare and quantify the overhead of the different modes of operation with regards to index size. 
%In this benchmark, the entire index is in-memory and we use synthesized 64-bit data points.
%\todo{Describe the considered synthesized workload and put that in perspective to stream (how many stream and what is the ingestion rate considered per stream.)}

%In the microbenchmarks, we consider the latency of index updates and query processing without accounting for network delay. 

%For the mhealth, we use data from Biovotion~\cite{medicalhealthtracker}, a health monitoring wearable which collects 12 different metrics at 50~Hz, where we set the chunk length to 10~s, i.e., up to 500 data points.

% 2428978 ingest/s  /  50 Hz = 48579.5675

\begin{figure}[t]
	\centering
	\includegraphics[width=\columnwidth]{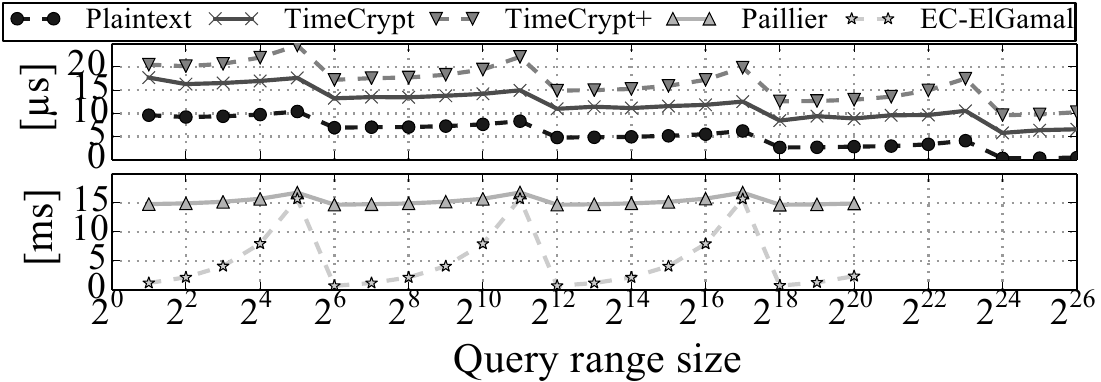}
	\vspace{-15pt}
	\caption{
		\setsmallcaption
		Aggregate queries over varying time ranges (i.e., query range size).
		Aggregating the entire index corresponds to retrieving the encrypted root. %\todo{why ASHE is not included here?}
		%For Paillier and EC-ElGamal the index size is capped at $2^{20}$ due to excessive overhead.
		%(k=64)
	}
	\lfig{eval:micro:latency}
	\vspace{-5pt}
\end{figure}

\subsecspacingtop
\subsection{Encrypted Data Processing Performance}
\subsecspacingbot

%\subsubsection{Index performance}
\lsec{eval:index}
We now discuss the evaluation results of different aspects of the encrypted index, as summarized in \rtab{complete-eval}.
In the microbenchmark, the index supports one statistical operation (i.e., sum) for isolated overhead quantification,
whereas in the E2E benchmark the index supports all our default queries.
In all experiments, we instantiate $64$-ary index trees and a keystream with one~billion keys via the key derivation tree.

\fakeparagraph{Index Size Expansion.}
To improve query efficiency, in-memory time series databases aggressively seek to reduce 
storage footprint, to support a model where almost all recent data can be stored in memory. %\red{e.g., Facebook Gorilla, BTrDB, and Netflix Atlas.} %Hence, enable serving fast queries at scale.
When considering encryption for time series data, 
the degree of ciphertext expansion has a direct impact on the encrypted index storage footprint, hence impacting query efficiency. 
%\red{Index size expansion due to encryption results in less data being represented in-memory.}
%As system memory is limited and generally smaller than available data, it is crucial to the system performance to keep the index-size small.
%\blue{To not diminish any performance gains of in-memory processing. 
%We carefully design HEAC with this consideration, for instance, the time-encoding allows us to eradicate the need to store key identifiers needed with stream ciphers, hence, introduce ciphertext expansion.} %that diminish in-memory performance gains.  
\oursystem has no ciphertext expansion for 64-bit values, \oursystemPlus introduces a 128-bit expansion due to the \mac tag. 
The encryption schemes in prior work~\cite{cryptdb, monomi, talos} exhibit large ciphertext expansion, 
e.g., for one million chunks we experience 96x index size expansion with Paillier.
Hence, limiting the performance gains of in-memory processing and impacting query latency.  
ASHE~\cite{seabed} uses an encoding where the expansion depends on the order of aggregation.
With in-range aggregation this amounts to 12.5\% higher expansion compared to \oursystem.
% 72 bytes per ciphertext compared to 64 bytes in \oursystem.%, or 192 bytes with \oursystemPlus.

%\todo{what is the ciphertext expansion of ASHE?}

\begin{comment}
The implication of a large index size due to ciphertext expansion is that less data can be represented by the index, 
limiting the performance gains of in-memory index processing and slowing down query response time.
\end{comment}

%For instance, the ciphertext size of a 64-bit value is for Paillier 786~Bytes and 165~Bytes for EC-ElGamal.  
%In \oursystem, we have no ciphertext expansion for 64-bit values.
%\oursystem introduces a low ciphertext expansion for values smaller than 64-bit, as ciphertexts in \oursystem have a fixed size of 64-bit.
%Note that due to the \chunksum randomization in \oursystem, compression cannot be used as in non-encrypted systems to further reduce the memory requirements~\cite{BTrDB}.

\fakeparagraph{Ingest Time.}
On each ingest, i.e., insertion of a leaf node, statistical aggregates of ancestor nodes are updated.
In \oursystem, additions are as efficient as in \emph{plaintext}.
Hence, the average ingest time increases slightly due to the encryption cost; 1.3x for the large index. %outperforming prior work by three orders of magnitudes.
With verification the average ingest time increases by 3.2x due to the \mac overhead.
%However, we outperform the strawman constructions by several orders of magnitudes (i.e. over 3000x faster).
%\red{The costs for updates are relatively high for both EC-ElGamal and Paillier, %with an average of 43~ms and 42~ms respectively at 1~million \chunksum records, 
%i.e., more than 3500x slower than plaintext.}
%\todo{For ASHE, the cost of updates is Xx than the plaintext.}

\begin{comment}
With Paillier this is due to the high encryption cost, while in EC-ElGamal the cost of elliptic curve additions dominates the average ingest time.
\end{comment}

\begin{comment}
    
\begin{figure}[t]
    \center
    \includegraphics[width=1\columnwidth]{images/paper_keyreg_latency_treesize}
    \vspace{-10pt}
    \caption{
    \setsmallcaption
         Performance of different
    %SHA2, AES, and AES-NI as 
    hash functions for the hash tree (AES-NI as default).
    The computation of a single key corresponds to $log(n)$ hashes, with $n$ keys (i.e., leaf nodes).
%    \oursystem relies on AES-NI which shows the best performance.
%    Note that without caching, two of such key computations are required for decryption of aggregated ciphertexts.
%    Each experiment is repeated 10000 times.
    }
    \vspace{-10pt}
    \lfig{eval:key-generation}
\end{figure} 
\end{comment}

\begin{figure*}[t!]
    \def\myheight{3.4cm}
    \centering
    % first plot
    \hspace{-8pt} 
    \begin{subfigure}[t]{0.46\columnwidth}
        \includegraphics[width=0.95\columnwidth]{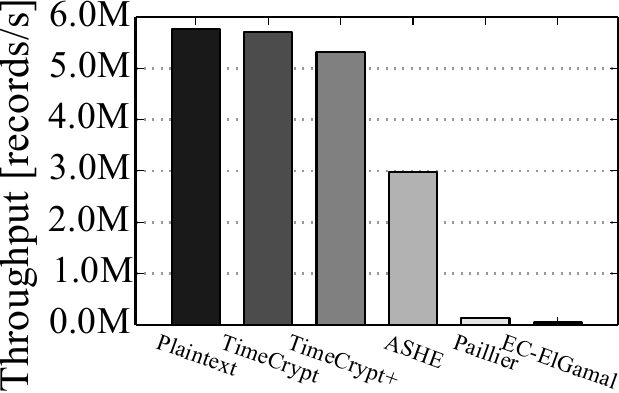}
        \caption{
            Ingests
        }
        \lfig{eval:e2e:tp:insert}
    \end{subfigure}
    % fourth plot
    \hspace{0pt} % space between the figures and captions
    \begin{subfigure}[t]{0.46\columnwidth}
        \includegraphics[width=0.95\columnwidth]{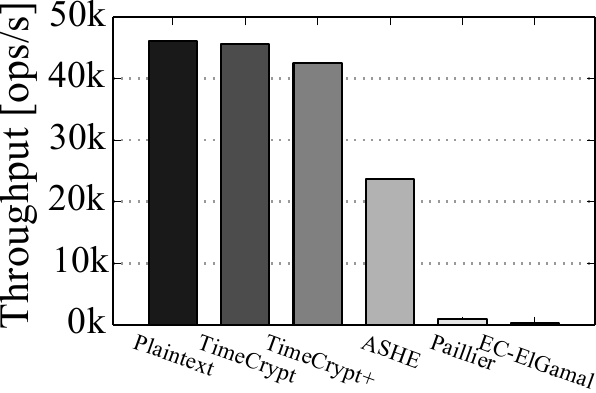}
        \caption{
            Statistical Queries
        }
        \lfig{eval:e2e:tp:querys}
    \end{subfigure}
    % third plot
    \hspace{5pt} % space between the figures and captions
    \begin{subfigure}[t]{0.47\columnwidth}
        \includegraphics[width=1.1\columnwidth]{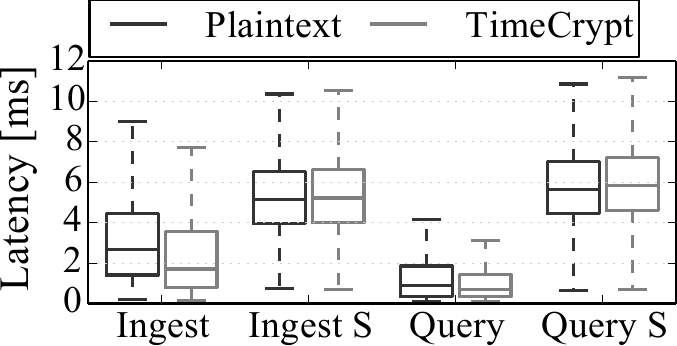}
        \caption{
            \oursystem vs. Plaintext
        }
        \lfig{eval:e2e:latency:ours}
    \end{subfigure}
    % second plot
    \hspace{10pt} % space between the figures and captions
    \begin{subfigure}[t]{0.6\columnwidth}
        \includegraphics[width=1.05\columnwidth]{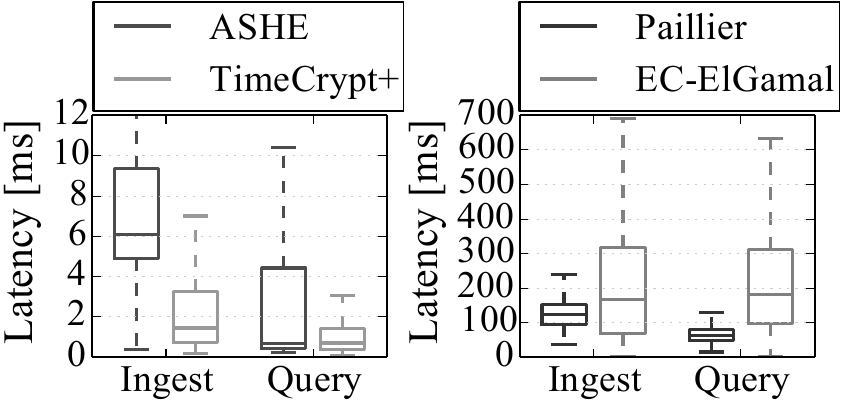}
        \caption{
          \oursystemPlus \& Alternatives
        }
        \lfig{eval:e2e:latency:strawman}
    \end{subfigure}
    \vspace{-8pt}
    \caption{
        \setsmallcaption
    Latency and throughput for ingest and statistical queries for \oursystem with HEAC vs. EC-ElGamal, Paillier, \& ASHE, and operating on plaintext indices.
        %The experiment is performed under 
        Heavy load experiment with a read-write ratio of 4 to 1, and also with extremely small~(S) index cache (1~MB).
        The AWS load generator creates 1200 streams with 100 clients, corresponding to 48579 streams in our health app 
        ($\Delta$:10s, 50Hz data rate).
        % 2428978 ingest/s  /  50 Hz = 48579.5675
%        (k=64)
     \lfig{eval:e2e}
    }
    \vspace{-8pt}
\end{figure*}

\fakeparagraph{Query Performance.}
\rfig{eval:micro:latency} shows the performance of the index for statistical range queries of different lengths, i.e., $[0,2^x]$ with $x \in [0..26]$.
As the length of queries increases fewer tree levels are traversed, which results in fewer cache fetches and lower computation time,
e.g.,  the index depth of five is observable in \rfig{eval:micro:latency}.
For plaintext and \oursystem the resulting pattern is similar due to the low cost of additions, while for \oursystem the decryption overhead is visible.
%{The prior work encryptions have higher addition costs, which results in the distinct sawtooth pattern due to on-the-fly aggregations within index nodes.}
%\red{EC-ElGamal has more expensive additions but faster decryptions than Paillier,
%which explains the similar performance with growing number of additions.} \todohs{remove}
Queries with non-power-of-k ranges require an index \textit{drill down} on either end of the range.
This increases the computation time logarithmic, $O(2(k$-$1)log_k(n))$ for a worst-case alignment, and not linear to the $n$ stored chunks.
%With a large $k$, the on-the-fly computations are higher. --> wrong, as the log_k becomes smaller and they balance each other out.
%Similar to ingest, \oursystem performs statistical queries with a latency close to the plaintext (i.e., only 1.1x). 
%and outperforms prior work significantly.

\fakeparagraph{Comparison to Alternatives.} 
%Thus far we have evaluated query and ingest performance of \oursystem relative to the plaintext performance. 
%\oursystem's performance is close to that of plaintext and can meet the performance needs of our target applications.
%\remove{To place TimeCrypt in context, we compare its processing performance on time-series workloads to existing encrypted databases, which vary in the type of functionalities they support and the type of workloads they are optimized for.} %that serve analytical queries and have been developed and evaluated at scale.
%These systems vary widely in many aspects, such as the type of functionalities they support on encrypted data and the type of workloads they are optimized for (e.g., batch analytics and transactional workloads). 
%For fairness, it is important to note that  our performance comparison focuses on one aspect, namely how these systems perform on time-series workloads. 
In \rfig{eval:e2e}, we show  HEAC's performance gains relative to the encryption schemes used in the other encrypted systems.
%We first evaluate HEAC's performance gains relative to the encryption schemes used in the other encrypted systems.
%We encrypt the digests with various schemes and compare the encrypted index performance. 
%\rfig{eval:e2e}~summarizes the results.
For this experiment we launch an ingest/query workload, with one machine and 100 threads, where each thread constantly performs four statistical queries after each chunk ingest.
The plaintext setting reaches a throughput of 5.77M records/s for ingest and 46.1k ops/s for statistical queries, as shown in~{Fig.~6a-b}. %\rfig{eval:e2e:tp:insert} and \rfig{eval:e2e:tp:querys}.
\oursystem demonstrates an outstanding throughput for both ingest and statistical queries with only 2.9\% slowdown compared to plaintext. 
With verification (\oursystemPlus), the slowdown increases to 7.8\% due to the larger index size and \mac computations.
\oursystem is by a factor of 2x, 20x, and 52x faster than ASHE, EC-ElGamal, and Paillier, respectively.
Despite ASHE's lower encryption and decryption cost, the system throughput is by 2x lower due to the higher aggregation costs on the server.
%(i.e., inserts and queries are more expensive depending on the number of aggregations).
This is due to ASHE's key-encoding updates, which \oursystem eliminates with the time-to-key mapping.
Fig.~6c-d shows the respective observed query latency.
The impact of a small index cache (1~MB) is distinct, but similar for both plaintext and \oursystem, due to higher cache misses. 
%With regards to latency, \oursystem outperforms EC-ElGamal and Paillier by two orders of magnitude and approaches 
%the latency of plaintext (Fig.~6c-d). 
%ASHE has a higher latency for inserts than \oursystem, but a similar query latency with more outliers due to the higher 
%aggregation costs.

\begin{comment}However, the impact of the cache is not visible in strawman, as the expensive crypto operations dominate.
Additions in EC-ElGamal saturate the CPUs, causing a high variance.
Paillier is rather bottlenecked at the client-side due to the expensive enc/decryptions (forcing us to scale up the Paillier client machine to 48 CPU cores).
\end{comment}

%\newpage
To compare how other encrypted systems perform while processing encrypted time series workloads,
we run one aggregate query over one billion data records on CryptDB, Pilatus, Seabed, and \oursystem.
Seabed requires seconds to process this query while CryptDB and Pilatus require minutes, 
whereas TimeCrypt can process such a query within a few milliseconds on a single machine.
\subsecspacingtop
\subsection{Client Performance}
\subsecspacingbot
\lsec{eval:client}

\begin{table}[t]
	\begin{center}
		\footnotesize
		\begin{tabular}[b]{m{1.0cm} l cccc}
			%		\toprule
			\hline
			& \multicolumn{2}{c}{HEAC} & \multicolumn{1}{c}{ASHE} & \multicolumn{1}{c}{Paillier}\\
			\cmidrule(lr){2-3}
			\cmidrule(lr){4-4}
			\cmidrule(lr){5-5}
			& [Enc/Dec]	& [\mac]  	& [Enc/Dec]	&[Enc/Dec] \\
			\midrule
			IoT 		&1.08ms& 20$\mu$s & 0.3ms & 1.59s / 1.62s\\ %530ms (128bit)
			%Phone	& \red{?} 	& \red{?}	& \red{?}	& \red{?} 	& 3.5ms	& 6.4ms \\
			Laptop	& 5.1$\mu$s 	& 0.2$\mu$s & 1.5$\mu$s / 1.3$\mu$s	& 30ms / 15ms \\
			% Cloud 	&6.98$\mu$s &6.98$\mu$s& 30~ms 	& 15~ms 	& 1.4~ms	& 1.1~ms\\
			%		\bottomrule
			%IoT 		&1.08ms& 1.1ms & 0.3ms/0.3ms & 1.59s/1.62s	& 252ms/N/A  \\ %530ms (128bit)
			%Laptop	& 5.1$\mu$s 	& 5.3$\mu$s & 1.5$\mu$s/1.3$\mu$s	& 30~ms/15ms 	& 1.4ms/1.1ms\\
			\hline
		\end{tabular}
	\end{center}
	% 18us * 30 *2
	\vspace{-10pt}
	\caption{
		\setsmallcaption
		Performance of crypto operations with at least 80-bit security and 32-bit integers on IoT devices (OpenMote) %with crypto hardware accelerator) 
		vs. laptop (MacBook).
		\oursystem uses a key derivation tree with $2^{30}$ keys.
	}
	\ltab{eval:crypto}
	\vspace{-10pt}
\end{table}

%\todo{maybe the prospective should as well include standard encryption schemes used by IoT for data protection, so as to help 
%the reader have a sense how much low this overhead.}
%\fakeparagraph{Crypto primitives.}
\rtab{eval:crypto} summarizes the enc/decryption and \mac costs of \enc in \oursystem.
%Though EC-ElGamal encryption is faster than Paillier, it exhibits a higher addition cost on the server.
\oursystem's cryptographic costs are dominated by the key derivation tree.
%By leveraging hardware AES instructions (AES-NI), 
%The derivation cost amounts to 2.5~$\mu$s for a tree with $2^{30}$ keys.
Enc/decryption amount to 5.1~$\mu$s, which accounts for the time to compute the key.
With \mac the clients incur 4\% higher costs. %another two AES operations are performed, requiring 0.15~$\mu$s for the additional KDF.
To put this in prospective, this is three orders of magnitude faster than Paillier, EC-ElGamal, and ABE schemes with only few attributes.
ASHE is faster in enc/decryption; the slight overhead in \enc is due to the cost of deriving keys from our key derivation construction to support access control. Though overall, \oursystem is more performant in ingest and query performance due to its faster aggregations.
%Nevertheless the overall cost of enc/decryption is still modest in \oursystem.
%The client-side overhead in \oursystem and ASHE grow with the supported query types.
%The computational overhead of the digest depends on the number and type of queries the system supports. 
%In the default configuration (i.e., three digests) with HoMAC the client side overhead is 16~$\mu$s for encryption and decryption.  %with a laptop and 3.3~ms on the IoT device. 
The overhead of resolution-based access is defined by the access granularity. 
For instance, with 10~s chunk intervals and minute and hourly resolutions, the encryption cost increases by only 1\% per day.

\fakeparagraph{Low Power Devices.}
\oursystem is particularly compelling for battery-powered constrained devices used in the IoT and environmental sensing, where the power consumption of encryption is a serious challenge.
Assuming one minute chunk intervals with \oursystem default queries, encryption consumes only 1.4\% (400mJ) more battery per day on an OpenMote device compared to sending data in the clear.

\subsecspacingtop
\subsection{Access Control} 
\subsecspacingbot
%\blue{The overhead of the crypto-based access control in \oursystem amount to the cost of:
%\textit{(i)}~access token sharing, 
%\textit{(ii)}~key derivation, and
%\textit{(iii)}~resolution-based access envelopes}
\shorten{\oursystem's access control mechanism differs from other encrypted data processing systems such as 
CryptDB~\cite{cryptdb}, Seabed~\cite{seabed}, and Pilatus~\cite{pilatus} in terms of granularity.
While these systems can support encrypted time series processing, limiting access within a time series is not possible, because 
the same key has to used for the whole stream to support computation.
Hence, to restrict access within a stream (e.g., only February), the accessible data has to be re-encrypted with a fresh key, which 
is then shared with the principal.}

%HEAC enables to specify varying access policies based on fine-grained keys on the same data stream. 
%HEAC enables the same data stream to be accessed under varying access control policies to support various degrees of protection. 
%This gives \oursystem the flexibility to support access policies with a varying granularity of protection for principals who are not known at the encryption time and with that, avoid any re-encryption costs for new policies.
%with varying granularity of protection levels.  
%Hence, an encryption-based access control that does not ensure re-encryptions costs towards new policies not known in at the 
%encryption time. 
In the following, we look at the performance and scalability of our encryption-based access control mechanism. 
The overhead can be quantified as the cost of key distribution, deriving \mac and 
encryption keys, and computing the resolution envelopes.
To characterize the overhead, we consider an example scenario where a data owner has 1000 streams and shares a subset of each stream with a data consumer.
%The overhead of envelopes in resolution-based access, and the computation cost of deriving the data decryption keys from these tokens.
%2xlog(n) 
%The transmission size of the tokens is 660 bytes for $2^{30}$ keys (i.e., per token 16 bytes key and 6 bytes tree encoding). 
\fakeparagraph{Na\"ive Key Management.}  
%\red{An access token encapsulates the nodes from the key derivation tree necessary to grant a principal access to a defined stream range. 
%The worst case number of shared nodes from the key derivation tree within an access token is $O(log(n))$, with $n$ keys. } 
\oursystem realizes access control by encrypting units of stream data with unique keys. 
Consequently, efficient key distribution is important for the scalability of this approach. 
In a  na\"ive approach, data owner can compile all the keys associated with the specified access policy and distribute the keys encrypted individually to each principle. 
However, this leads to access tokens of size $O(n)$ where $n$ is the number of keys (i.e., units of stream data included in the access policy).
With our key derivation construction, we have a logarithmic worst-case complexity in the number of shared stream units $O(log(n))$.

\fakeparagraph{Communication.}  
An access token in \oursystem contains in the worst-case $2(\log(n) - 1)$ inner nodes of the tree key-derivation construction where $n$ is the number of keys in the tree. % 22~byte nodes
This reduces the communication cost from a na\"ive approach from 50~GB to 1.28~MB, considering one year of data shared in our example scenario.
With resolution-based access policies, the data consumer has to additionally download two envelopes per aggregation query (72~additional bytes).

\begin{comment}
The communication costs consist of the size of 1000 access tokens, where a token contains key derivation tree nodes covering the shared interval. 
The tree node size is 22 bytes (i.e., 6 bytes key and 6 bytes tree encoding) and the number of tree nodes per token is between 1 and $2(\log(n) - 1)$ where $n$ is the number of keys in the tree.
This amounts to 38 nodes (0.84kB) with $2^{20}$ keys, and 58 nodes (1.3kB) with $2^{30}$ keys.
Hence, the state to share for the stream owner with the principal is between 1~kB and 1.28~MB.
If the access token restricts the resolution, the principal has to download two envelopes per aggregation query of a stream, which amounts to 72~bytes additional bytes (7.2 kB for a query over 1000 streams).
\end{comment}

%The communication cost can be broken in two parts: (1) Access Tokens: this cost in encountered once at the time of creating access policy and consists of exchanging 1 token with the client that encapsulate a maximum of O(log(n)) intermediate noes of our key derivation constructions. 
%(2) Envelops:  Each query result comes accompanied with two envelops. The cost of this is X. 

\fakeparagraph{Computation.}  
%The data owner computes the tree nodes for each of the 1000 tokens.
%The computation time for deriving a tree node or a key without cashing has an upper bound of $log(n)$, i.e., 2.5~$\mu$s. %for a tree with $2^{30}$ keys. 
Deriving the access token for all streams requires 145~ms.
The decryption keys can be computed at a rate of 400k per second.
With resolution-based access, the principal has to perform an additional decryption (for the envelope), which reduces the rate to 380k keys per second.

\fakeparagraph{Storage.} 
The storage cost can be broken down into two parts;
key storage at the data consumer (1.28~MB), and resolution-related keying material at the server, which
grows linearly with time (i.e., the envelopes).
%Key storage at the principal is the size of the token for each stream, discussed in the communication cost with 1.28~MB worst case for 1000 streams.
%On the server, each envelope adds storage overhead of 36 bytes.
With a stream that consists of 10~s chunk intervals over one year with hour/day/month resolution support, the server stores  1.6~MB keying material (45.7k envelopes) per stream.

%The data owner maintains 1 master tree node per stream (i.e., 2.2 kB secret material for 1000 streams).}
%\todo{ The communication cost can be broken in two parts: (1) key storage at the principals and (2) the storage cost for the envelopes needed for resolution}. 
%\fakeparagraph{Comparison to alternatives.}   

\begin{comment}
The storage cost can be broken in two parts, key storage at the principal, the storage cost for the envelopes on the server, and the master key storage at the owner. 
Key storage at the principal is the size of the token for each stream, discussed in the communication cost with 1.28~MB worst case for 1000 streams.
On the server, each envelope adds storage overhead of 36 bytes.
With a stream that consists of 5s chunks over one year with hour/day/month resolution support, the server has to store  1.6~MB data with 45.7k envelopes.
For 1000 streams, the server stores 1.6~GB of envelopes.
The data owner maintains 1 master tree node per stream (i.e., 2.2 kB secret material for 1000 streams).
\end{comment}

\begin{figure}[t]
    \center
    \includegraphics[width=0.95\columnwidth]{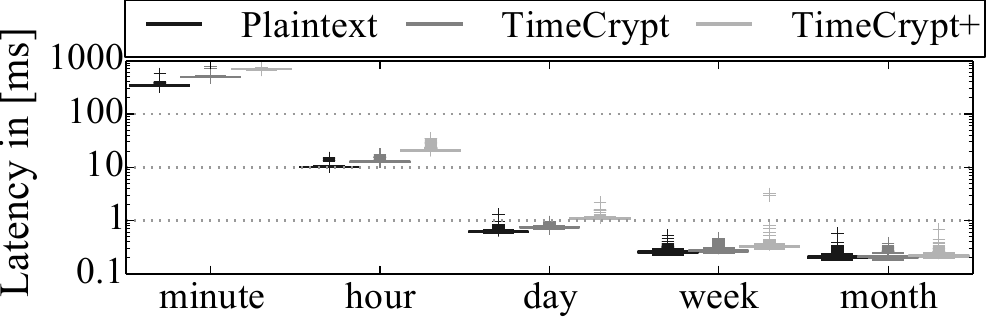}
    \vspace{-8pt}
    \caption{
            \setsmallcaption
            Latency in log-scale for statistical queries of one month data in our health app (121M records). 
            The x-axis shows the granularity of the requested data from one minute to one month.
            %We compare \oursystem to operating on clear data.
            %The higher the number of decryptions, the larger the overhead of \oursystem.
            %E.g., with one minute resolution \oursystem's latency is by 1.51x higher. %, which is due to the high number of decryptions. 
            %The overhead decreases to only 1.01x for the one month resolution. 
            %[ 321.348709    9.543899    0.99014     0.684539    0.638347]
            %[ 484.6573035   12.656946     1.1306925    0.708007     0.6479435]
            %TimeCrypt Overhead: 1.51x    1.32x   1.14x    1.03x    1.01x
    }
    \lfig{eval:e2estat} 
   \vspace{-8pt}
\end{figure} 
  
\fakeparagraph{Comparison to an ABE-based Approach.}
Although, key-policy attribute-based encryption (KP-ABE) (used in Sieve~\cite{Sieve}) is a powerful tool for access control, it comes with a relatively high computational cost, especially for low-power devices and 
when used to enable fine-grained polices as needed in time series data.
Compared to KP-ABE (implementation from~\cite{abe-library-frauenhofer}), \enc is three orders of magnitude more efficient for encryption/decryption.
%\remove{If an IoT device encrypts one chunk per minute, an ABE-based solution cuts the battery life by one order of magnitude compared to \enc.}\todo{Lukas, can we rephrase this to:}
For an IoT device encrypting one chunk per minute, an ABE-based solution drains one order of magnitude more battery life compared to \enc.
Additionally, ABE does not support computation on encrypted data.
%\remove{While KP-ABE can be employed with hybrid encryption to increase efficiency}~\cite{Sieve},
%\remove{this solution does not allow for resolution-based access control, and reduces the fine-granularity of the access control.}} 
\fakeparagraph{Interrupt Key Canceling.}
\oursystem can add epoch borders to reduce the risk of leakage from aggregating the skipped interval between two shared non-continuous intervals
%that interrupt the key canceling during aggregation to reduce the risk of revealing sums between multiple shared intervals of a stream 
(\rsec{crypto:primitives}).
Each additional epoch border within the query range incurs an additional computational cost to decryption (i.e., one key derivation and two additions). 
For example, considering a weekly epoch and a daily epoch in a data stream, the decryption cost for a monthly aggregate result increases by a factor of 2.5x and 14.5x, respectively. 
%from $5.1~\mu$s to $12.8~\mu$s and $74~\mu$s, respectively. 
However, even for fine-grained epochs (e.g., over 300 per range), the decryption latency remains well below 1~ms and would not impact user perception.

\subsecspacingtop
\subsection{Applications} 
\subsecspacingbot
\lsec{eval:e2e}
%\todo{Shall we call this section Applications?}
In this section, we evaluate the end-to-end overhead of \oursystem and its effectiveness in running complex, real-world applications.
We developed four apps atop of \oursystem that represent different challenging requirements and workloads.     

\fakeparagraph{mHealth Views - Interactivity.}
We implemented an mHealth dashboard for the Biovotion health tracker~\cite{medicalhealthtracker}.
The dashboard shows summary plots of the underlying data (i.e, windowed AVG).
% such as HeartRate, Skin Temperature, Steps etc., which mainly queries summarizations over the raw data (i.e, windowed AVG).
The data consists of 12 different metrics at 50~Hz from the Biovotion sensor over two weeks, which we stretch to one year worth of data.
\rfig{eval:e2estat} shows the response time for aggregation plots of last month's data (121M records).
We also consider the extreme case of plotting one-month data at minute granularity (403~MB plot), which induces an overhead 
of 1.45x (2.0x for \oursystemPlus) in latency compared to plaintext.
%This is due to the high number of decryptions of the individually retrieved aggregates (i.e., 40320).
With lower granularity, the overhead sharply decreases and reaches 1.06x (1.29x for \oursystemPlus).

\begin{comment}
\fakeparagraph{mHealth Views - Interactivity.}
\blue{We back a mHealth dashboard application for Biovotion~\cite{medicalhealthtracker} health tracker with \oursystem. 
The dashboard shows summary plots of the underlying raw data such as HeartRate, Skin Temperature, Steps etc. 
The goals is to show application developed with \oursystem can meet low-latency requirements dashboard, which performs summarization queries over the raw data (i.e, windowed AVG).}
\blue{\fakeparagraph{Data.} 
The data consists of 12 different metrics at 50~Hz from the Biovotion sensor over two weeks, and we stretch the data to one year.}
\blue{
\fakeparagraph{Results.} 
In \rfig{eval:e2estat}, we show the response time for aggregation plots of last month's data (121M records) for a specific source.
We also consider the extreme case of plotting one-month data at minute granularity (403~MB plot), which induces an overhead of 1.45x (2.0x for \oursystemPlus) in latency compared to plaintext.
This is due to the high number of decryptions of the individually retrieved aggregates (i.e., 40320).
With higher granularity, the overhead sharply decreases and reaches 1.06x (1.29x for \oursystemPlus) for one month.}
\end{comment}

\fakeparagraph{DevOps Trend Detection - Complex Analytics.}  
We developed a trend detection app for CPU utilization.
We use a CPU monitoring dataset generated by the time series benchmark suite~\cite{timescalebenchmarkframework} with 10 metrics, 10s data rate, and per minute chunk size $\Delta$ over one year.
The results of a two-dimensional linear regression model on different ranges of an encrypted CPU monitoring stream are shown in \rfig{eval:ml}.
\shorten{With a similar setup as in \rfig{eval:e2e} for both plaintext and \oursystem, we observe an ingest throughput of 60.6k records/s (corresponds to approx. a load of 600k machines) and a query throughput of 40.4k ops/s.}
\oursystem matches the plaintext performance (0.75\% slowdown).

\fakeparagraph{Smart Energy Service - Access Control Scalability.}  
We extended a smart meter application, where a service computes the aggregated energy consumption per day over households.
Each smart meter uploads a chunk every 5s, but the service can only compute per day aggregates for each stream. 
We use the ECO dataset~\cite{kleiminger-eco-dataset}, which contains smart meter data sampled at 1~HZ rate and collected over 8 months.
\rfig{eval:multi:stream} shows the query latency for the aggregated energy consumption over up to 1000 streams.
\oursystem's overhead is attributed to multi-stream processing and resolution-based access.
The overhead stems from the linearly increasing decryption costs in the number of streams that are aggregated.

\fakeparagraph{Crowdsourced mHealth - Privacy Policy Transformation.} 
We enhance the mHealth app with a crowdsourcing feature which enables users to opt-in their data to be part of crowdsourcing for a targeted research project, as described in \rsec{extensibility}.
For $n$ users, the secure aggregation protocol~\cite{bonawitz2017practical} adds a communication overhead of $n$ Diffie-Hellman key exchanges per user to create the envelopes.
The envelope enc/decryption increases linearly (e.g., below 1~ms for 100 users).

\begin{figure}[t]
	\centering
	% first plot
	\begin{subfigure}[t]{0.49\columnwidth}
		\includegraphics[width=\columnwidth]{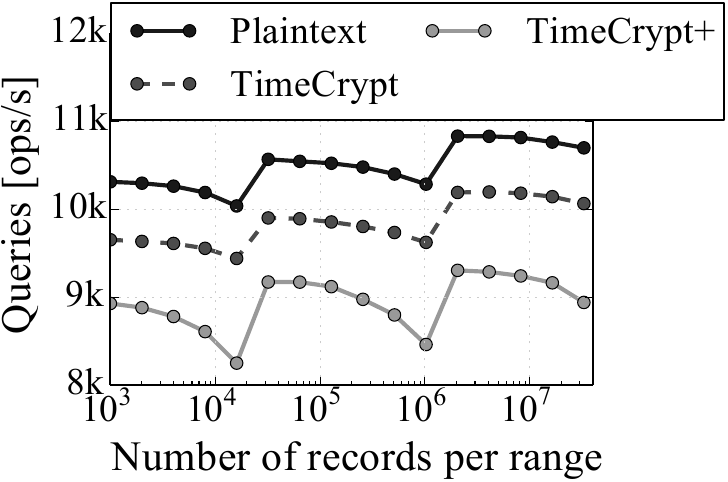}
		\caption{
			\setsmallcaption
			DevOps Application
		}
		\lfig{eval:ml}
	\end{subfigure}
	% fourth plot
	\hspace{0pt} % space between the figures and captions
	\begin{subfigure}[t]{0.48\columnwidth}
		\includegraphics[width=\columnwidth]{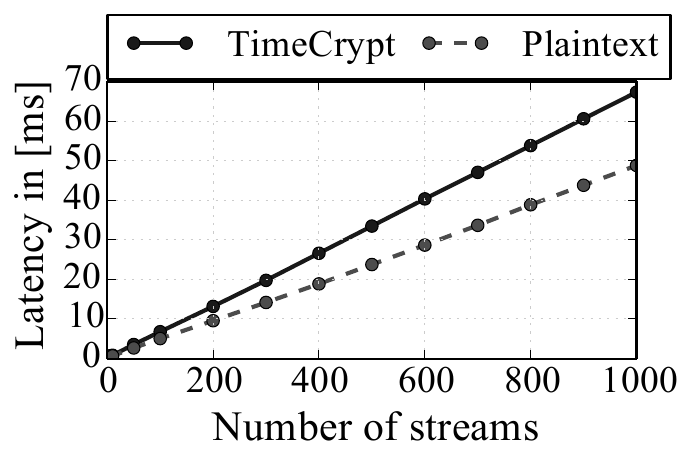}
		\caption{
			\setsmallcaption
			SmartEnergy Application
		}
		\lfig{eval:multi:stream}
	\end{subfigure}
	\caption{
		\setsmallcaption
		Applications: (a) DevOps trend detection queries on CPU utilization over different number of records. (b) Energy consumption queries for a day over multiple streams in a smart meter application.
	}
	\vspace{-15pt}
	\lfig{eval:applications}
\end{figure}

\secspacingtop
\section{Related Work}
\secspacingbot
\lsec{related-work}

There is a large body of research on privacy-preserving systems, encrypted search, and secure outsourced computation. For brevity\blue{,} we focus our discussion here on works that are closest to TimeCrypt. 

\fakeparagraph{Encrypted Databases.}
%Several encrypted database systems~\cite{cryptdb, monomi, talos, pilatus, seabed} have been introduced in the past years.
%Established and startup companies have attempted to adopt encrypted databases~\blue{\cite{}}.
%Currently, there is no one solution for all use cases, and system designers need to understand their system's
%and workloads' characteristics.
Fuller et al.~\cite{fuller2017sok-EDB} provide a comprehensive overview of the encrypted database landscape.
We now discuss several works in this space that are analogues to \oursystem.
CryptDB~\cite{cryptdb} and Monomi~\cite{monomi} augment relational databases with encrypted data processing capabilities, 
however, encryption schemes used in these systems are not efficient enough to support interactive queries on large data.
Seabed~\cite{seabed} focuses on Spark-like batch processing workloads and resorts to symmetric partial-homomorphic 
encryption to enable interactive queries on big data but without the tight latency requirements of time series data.
CryptDB, Monomi, and Seabed do not support cryptographic access control or verifiable computation, as the case with \oursystem.
 %are designed mainly for transactional~\cite{cryptdb, pilatus, monomi, enki} and Spark-like batch processing workloads~\cite{seabed} without the tight latency requirements of time series data.
%Moreover, they operate within a single-user model, i.e., do not provide crypto-enforced fine-grained access control that is our focus.
%Hence, they neither provide the necessary primitives for stream analytics nor support crypto-enforced fine-grained access.
%E.g., CryptDB~\cite{cryptdb} has high computation load and large memory expansion. 
%Moreover, it relies on a trusted proxy to handle sharing and process queries.
ENKI~\cite{enki} and Pilatus~\cite{pilatus} support sharing and encrypted computations
but they scale poorly with the number of principals and the size of data. Also, they do not support fine-grained policies.
Bolt~\cite{bolt} is an encrypted data storage system for time series data that supports retrieval of encrypted chunks but does not support server-side computation on encrypted data or fine-grained sharing.
%However, Bolt does not support statistical range queries on encrypted data nor crypto-enforced fine-grained access.
%Bolt is primary a storage service 
BlindSeer~\cite{blindseer} enables private boolean search queries over an encrypted database by building an index with Yao's garbled circuits and primarily targets private search over large data with no support for statistical queries.
It integrates access control for search queries
but requires two non-colluding parties.
Another line of research considers building data processing systems in trusted execution environments~\cite{haven, vc3, enclavedb}, which can provide confidentiality and integrity of queries.
%However, trusted hardware based system require dedicated hardware and their trust model implies considering the TEE 
%trustworthy, which may be vulnerable to attacks~\cite{sgxcacheattack, sgxpectre}. 
In \oursystem, we do not require dedicated hardware and rely on cryptographic primitives to ensure confidentiality and integrity of computation.

\fakeparagraph{Cryptography-based Access.}
%\todo{Extreme care to be factual and cover more work!}
%on untrusted servers 
Cryptographically enforced access control is explored by crypto-systems \cite{abe5} such as
%(hierarchical) 
identity-based encryption, attribute-based encryption (ABE), predicate encryption, and functional encryption. 
They enable complex access control to encrypted data. 
ABE~\cite{abe1, abe2, abe3, abe4, goyal2006attribute, agrawal2017fame} is the most expressive among them, though
it comes with limitations with respect to fine-grained access and and dynamic updates~\cite{abe5}.
%Recent ABE schemes, such as FAME~\cite{agrawal2017fame}, which exhibit a constant decryption time (60 ms) independent of the number of attributes, still suffer from linear increase during encryption which is critical for the large volume and continuous data types.
%\red{Sieve~\cite{Sieve} combines ABE with key-homomorphic encryption, to enable in situ re-encryption of user's data  without 
%revealing new keys to the cloud.}
Current ABE-based systems lack homomorphic capabilities (i.e., no computation on ciphertexts) and scalability required for time 
series data workloads.
In general, adding homomorphic capabilities to ABE remains an open challenge~\cite{HABE3}.
Recently, important progress has been made on constructions of homomorphic attribute based 
encryption~\cite{HABE1,HABE3,HABE2,HABE4}. However, they remain limited in functionality and are computationally expensive.
A related line of work is searching over encrypted data with predicate evaluation~\cite{shi, boneh2007conjunctive}.
While predicate encryption schemes~\cite{shi, boneh2007conjunctive} support range queries over encrypted data, they lack the 
required efficiency in our setting, as they require a linear scan through the database and also due to their underlying computationally expensive pairing-crypto. 
\secspacingtop
\section{Conclusion}
\secspacingbot
\lsec{conclusion}
%\vspace{-0.5em}

%\oursystem is a new system that augments time series data stores with support for encrypted data processing.
In this paper, we presented \oursystem, a new scalable system that enables fast analytics over large encrypted data streams.
\oursystem introduces \enc, a novel encryption construction
that enables execution of real-time analytics over encrypted stream data and empowers data owners to enforce access restrictions on encrypted data based on their privacy and access control preferences. 
%We design and build \oursystem.
Our evaluation on various large-scale workloads shows 
\oursystem's performance is close to operating on plaintext data,
demonstrating the feasibility of providing high-performance and strong confidentiality guarantees when operating on 
large-scale sensitive time series data.

%\section*{Availability}
%\todo{no need for this just add the link as footnote to third item of contributions in the intro.}
%An open-source implementation of \oursystem is available at \url{https://github.com/TimeCrypt/timecrypt}.
\vspace{-5pt}
\section*{Acknowledgments}
%\todo{TODO}

%We thank  Amy Ousterhout, Scott Shenker, and Friedemann Mattern for their detailed feedback on 
%earlier versions of this paper. \blue{We also thank our anonymous reviewers for their valuable feedback, and our shepherd Ben Zhao
%who helped shape the final version of this paper.}

We thank our shepherd Ben Zhao, the anonymous reviewers, Amy Ousterhout, Scott Shenker, and Friedemann Mattern for their valuable feedback. This work was supported in part by the Swiss National Science Foundation Ambizione Grant, VMmware, Intel, and the National Science Foundation under Grant No.1553747.

%\bibliographystyle{IEEEtran}
%\bibliography{IEEEabrv,biblio}
%\bibliographystyle{ACM-Reference-Format}
%\bibliography{biblio}
\bibliographystyle{plain}
\bibliography{biblio}

%\appendix
\newpage
% !TEX root = ../paper_secure_TS.tex
\section{Appendix}
\lsec{appendix}
\subsection{\oursystem Encryption}
\lsec{proof}

In this section, we analyze and proof the security of \oursystem's cryptographic construction for encrypting the chunk \chunksum.
We first outline the basic cryptographic building blocks of our construction and give a detailed definition of our encryption scheme. 
Then, after analyzing the tree construction of \oursystem, we provide a proof of security of the proposed scheme.

\subsubsection{Used Cryptographic Building Blocks}
\lsec{appendix:cryptodefinitions}

In our construction, we make use of the following cryptographic primitives.

\fakeparagraph{Pseudorandom Function (PRF).} 
A function $F: \{0,1\}^\lambda \times \{0,1\}^n \rightarrow \{0,1\}^m$ is a PRF, if there is no probabilistic polynomial-time (PTT) distinguisher, which can distinguish $F_k(x) = F (k, x)$ from a random function drawn from $\{f: \{0,1\}^n \rightarrow \{0,1\}^m\}$ with non-negligible probability in $\lambda$ where $k$ is drawn uniformly at random from $\{0,1\}^\lambda$~\cite{goldreichconstuction}. 

\fakeparagraph{Pseudorandom Generator (PRG).} 
$G: \{0,1\}^n \rightarrow \{0,1\}^m$ is a pseudorandom generator, if $m>n$ and no probabilistic polynomial-time (PTT) distinguisher can distinguish the output $G(x)$ from a uniform choice $r \in \{0,1\}^m$ with non-negligible probability~\cite{goldreichconstuction}.

\fakeparagraph{Goldreich-Goldwasser-Micali Construction.} 
The Goldreich-Goldwasser-Micali construction shows how to construct a PRF from pseudorandom generators~\cite{goldreichconstuction}. 
Given a PRG $G: \{0,1\}^s \rightarrow \{0,1\}^{2s}$ and $G(x) = G_0(x)||G_1(x)$ both of length $s$, a PRF $F: \{0,1\}^\lambda \times \{0,1\}^n \rightarrow \{0,1\}^s$ can be constructed as
%\begin{equation}
$F(k, x=x_1, x_2...x_n) = G_{x_n}(...(G_{x_2}(G_{x_1}(k))))$.
%\end{equation}
Given $G$ is a pseudorandom generator then the above construction is a pseudorandom function.

\fakeparagraph{Constrained Pseudorandom Functions (PRF).} 
A PRF  $F: \{0,1\}^\lambda \times \{0,1\}^n \rightarrow \{0,1\}^m$ is constrained with respect to $S \subseteq \{0,1\}^n$, if it supports two additional algorithms \textit{F.constrain} and \textit{F.eval} and has an additional key space $\mathfrak{K}_c$~\cite{boneh2013constrained}.

\textit{F.constrain($k$, $S$)}: On input $k \in \{0,1\}^\lambda$ and set $S$, the algorithm computes a key $k_S \in \mathfrak{K}_c$. $k_S$ enables to evaluate $F(k,x)$ for all $x \in S$ but no other value. 

\textit{F.eval($k_S$, $x$)}: On input $k_S \in \mathfrak{K}_c$ (constrained key) and $x \in \{0,1\}^n$, the algorithm evaluates $F(k,x)$ for $x$ if $x \in S$ else outputs $\perp$.

\subsubsection{Scheme Definition}
\lsec{security-of-encryption-scheme}
\oursystem introduces a new variant of the Castelluccia encryption scheme~\cite{castelluccia2005, castelluccia2009}, which in addition to additive homomorphic computations also allows for access control. 
The basic idea of the Castelluccia encryption scheme is to replace the exclusive-OR operation in a standard stream cipher with modular addition. 
We leverage the same principle, but extend it with a key derivation function based on a binary tree (i.e., for access control) and an encoding for reducing the number of keys required for decryption on in-range aggregated ciphertexts.
Let $TreeKD(k, t): \{0,1\}^\lambda \times \{0,1\}^n \rightarrow \{0,1\}^\lambda$ be our key derivation function with master secret $k$ computing the $t$-th key, we define our symmetric private-key encryption scheme $(Gen,Enc,Dec)$ as:

%\begin{itemize}
 %   \setlength\itemsep{1em}
\fakeparagraph{Gen}: on input $1^\lambda$, randomly pick $k \in \{0,1\}^\lambda$ for the key derivation function and set the plaintext space to $[0, M-1]$ where $M = 2^\lambda$.
\fakeparagraph{Enc}: on input $m \in [0, M-1]$, samples $t \in \{0,1\}^n$  uniformly at random and encrypts message $m$ as $c := \langle m + TreeKD(k, t) - TreeKD(k, t+1) \bmod M, t\rangle$.
\fakeparagraph{Dec}: on input $c'$, decrypts ciphertext $c'= \langle c, t\rangle$  as $m = c - TreeKD(k, t) + TreeKD(k, t+1) \bmod M$.
%\end{itemize} 

We observe that the above scheme is additively homomorphic if we expand the added ciphertext with the used parameters $t$ during encryption.
Note that for the analysis, we select the value $t$ uniformly random and attach it to the ciphertext.
However, in our practical system, we do not attach each parameter $t$ to the ciphertext and can select $t$ based on the time-counter without compromising the security.
As long as the selected values for $t$ are unique (i.e., do not repeat) the security guarantees remain intact. 
%Furthermore, we can reduce the number of key derivations for decryption on in-range aggregated ciphertexts by canceling out the keys in between.

To prove the CPA-security of our scheme, we first analyze our key derivation function based on a tree data structure and show that the function is similar to a pseudorandom function.
In a second step, we show that our encoding in the encryption step, which requires two evaluations of a pseudorandom function, can be reduced to a single pseudorandom function. 
Finally, we proof the CPA-security on the simplified scheme, which is similar to the scheme analyzed in~\cite{castelluccia2009}.

\subsubsection{\oursystem Tree}
\lsec{prooftree}
In our system, we use a tree-based key derivation function $TreeKD$, which allows for access control.   
%We use a similar construction as presented in~\cite{treekeyreg}. 
We define the function $TreeKD$, which derives keys based on a binary tree with height $h$.
Keys are derived from the $2^h$ leaf-nodes of the tree. Each node in the tree has a unique label $l$ in $\{0,1\}^*$ and an associated \textit{tree-key} in $\{0,1\}^\lambda$.
%Leaf-nodes additionally contain an \textit{external-key} in $\{0,1\}^\lambda$, which corresponds to an output value of the key-derivation function. 
We define the label of each node in the following manner. The root node of the tree has the label $\epsilon$, the empty-string, 
whereas the left and right children of a node with label $l$ have the labels $l||0$ and $l||1$ respectively. 
Hence, the leaf nodes $ln$ are indexed by their label $x$ where $x \in \{0,1\}^h$. 
We denote a leaf node with label $x$ as $ln_{x}$.

Each node in the tree with label $l$ has a \textit{tree-key} $z_l$
The root-node of the tree has a randomly chosen \textit{tree-key} $z_\epsilon$ in $\{0,1\}^\lambda$, which corresponds to the input key of the key-derivation function.
To derive the keys for the children of a node, a pseudorandom generator $G : \{0,1\}^\lambda \rightarrow \{0,1\}^{2\lambda}$ is used. 
Let $G_0$ and $G_1$ be defined as $G(k) = G_0(k)||G_1(k)$, where $|G_0(k)| = |G_1(k)| = |\lambda|$ and $k \in \{0,1\}^\lambda$. 
The left and right child of a node with \textit{tree-key} $z_l$ are computed as $z_{l||0}=G_0(z_l)$ and $z_{l||1}=G_1(z_l)$.
 Hence, the \textit{tree-key} $z_l$ of a leaf-node $ln_{l}$ is constructed as 
 %\begin{equation}
%\lequ{eq:keyderleaf}
 $z_l=G_{l_{h}}(...(G_{l_2}(G_{l_1}(z_{\epsilon}))))$
 %\end{equation}
 Note that if a \textit{tree-key} $z_l$ is revealed, it is easy to compute the \textit{tree-keys} of its children, but two children \textit{tree-keys} do not reveal any information about the parent \textit{tree-key}.
This property allows for access control.%~\cite{treekeyreg}.
 
 To derive the key for input value $t$, the function $TreeKD(k, t)$ computes the \textit{tree-key} $z_t$ of the leaf-node $ln_t$ with root-node key $z_\epsilon = k$ and outputs $z_t$.
 Hence, the function $TreeKD(k, t): \{0,1\}^\lambda \times \{0,1\}^h \rightarrow \{0,1\}^\lambda$ is defined as
 \begin{equation}
    \lequ{eq:keyder}
    TreeKD(k, t) = G_{t_{h}}(...(G_{t_2}(G_{t_1}(k))))
\end{equation}
where $t=t_1,t_2,..,t_{h}$. \newline

 \fakeparagraph{Claim 1.} \textit{$TreeKD$ is a constrained pseudorandom function.}
 
 \textit{Proof.} This directly follows from the definition of the Goldreich-Goldwasser-Micali construction because $TreeKD$ has an identical definition.
 Furthermore, $TreeKD$ can be constrained by sharing only inner nodes~\cite{boneh2013constrained}.
  \newline
 \subsubsection{Encryption Scheme Proof Sketch}
 \lsec{secproof}
 \fakeparagraph{Lemma 1.} \textit{Given any sequence of $q$ distinct $n$-bit values $x=(x_1,...,x_q), y=(y_1,...,y_q)$ where $q < 2^\lambda$, then there are at least $(2^\lambda)^{2^\lambda-q}$} functions $f$ such that $F(x_i)=y_i$ for all $i$, where $F=f(x)-f(x+1)$.

 \textit{Proof.} We use a similar proof construction as presented in~\cite{prfaddproof}. 
 Let $X=\{x_1,...,x_q\}$ and $S= \{0,1\}^\lambda \setminus X$ be the selected sets and $S$ non-empty.
 We show how to enumerate all functions $f$.
 For all $s \in S$ we can set the function result of $f$ to any $\lambda$-bit string independently,
 whereas the remaining values $x \in X$ can be computed as $f(x) = f(s) + F(s-1) + F(s-2) + ... + F(x)$ where $s$ is the smallest $\lambda$-bit string such that $s > x$.
 Note that the addition and the comparison are all modulo $2^\lambda$ and $s-1, s-2, ..., x \in X$ (i.e., s is the smallest $\lambda$-bit string, which is greater than $x$).
With this construction of $f$, we can verify that $F(x_i) = y_i$ for all $i$ since each intermediate result cancels out.
Since $|S| = 2^\lambda - q$, we first have $2^\lambda - q$ choices for the argument of the function $f$ and for each choice $2^\lambda$ possible results, which accumulates to total of $(2^\lambda)^{2^\lambda-q}$ possible functions.
\newline

\fakeparagraph{Collary 1.} \textit{Let $O$ and $r$ be uniform random functions on $\{0,1\}^\lambda \rightarrow \{0,1\}^\lambda$ and define function $R : \{0,1\}^\lambda \rightarrow \{0,1\}^\lambda$ as $R(t)=r(t) - r(t+1)$. 
For any oracle algorithm $A$ that is bound to $2^\lambda - 1$ queries to the oracle, the distribution of the algorithm for $O$ or $R$ as the oracle is the same.}

\textit{Proof.} This follows directly from Lemma 1, since the number of queries of $A$ is bound to $2^\lambda - 1$. 
\newline

%\{0,1\}^\lambda
 \fakeparagraph{Lemma 2.} \textit{If function $f:\{0,1\}^\lambda \times \{0,1\}^\lambda \rightarrow \{0,1\}^\lambda $ is a pseudorandom function and the corresponding distinguisher is bound to $q \leq 2^\lambda$ queries then the function $F: \{0,1\}^\lambda \times \{0,1\}^\lambda \rightarrow \{0,1\}^\lambda$ defined as  $F(k,t) = f(k,t) - f(k,t+1) \bmod 2^\lambda$ is a pseudorandom function with a distinguisher bound to $q/2$ queries.}

 \textit{Proof.} Let $A$ be a PPT adversary, which can distinguish $F$ from a random function $R : \{0,1\}^n \rightarrow \{0,1\}^\lambda$ with non-negligable probability with $q/2$ queries to the oracle. 
 We show with a proof by reduction that given $A$, we can construct a polynomial-time distinguisher $D$ emulating $A$, which can distinguish the PRF $f$ from a random function with non-negligible probability.
 $D$ has access to an oracle function $O: \{0,1\}^n \rightarrow \{0,1\}^\lambda$ and emulates $A$ to decide if $O$ is pseudorandom or random by outputting a bit $b \in \{0,1\}$. 
% We construct $D$ as follows.
Whenever $A$ performs query with value $t$, $D$ queries the oracle for $O(t)$, $O(t+1)$ and responds with $O(t) - O(t+1) \bmod 2^\lambda$.
If $A$ outputs the bit $b'$, the distinguisher $D$ outputs the bit $b=b'$. 
If $A$ runs in polynomial time also $D$ runs in polynomial time. 
We can observe that given attacker $A$ has a non-negligible advantage in distinguishing $F$ from random, the distinguisher $D$ can distinguish $f$ with the same probability, which follows from Lemma 1 and Collary 1.
%The intuition is that if $O$ is random then also the input to $A$ is random, and if $O$ is pseudorandom also the input to $A$ is pseudorandom. 
%Nevertheless, if the number of queries by attacker $A$ is bounded by a polynomial function $p(n)$, then $D$ requires $q(n) = 2p(n)$ queries to the oracle.   
The intuition is that a distinguisher distinguishing $F$ from a function $U(t) = R(t) - R(r+1)~mod~2^\lambda$  has the same distribution as a distinguisher distinguishing $F$ from $R$ as long as the distinguisher is bound to $2^\lambda - 1$ queries.
The distinguisher $D$ requires $q$ queries to the oracle if $A$ is bounded by $q/2$ queries.
Hence, if $A$ has an advantage $\epsilon$, $D$ has an advantage $\epsilon$ in distinguishing $f$ from $r$. 
$F$ is an aggregate pseudorandom function. For the formal definition of an aggregate PRF, we refer to~\cite{Aggregatable-prf}.

With Claim 1 and Lemma 2, we can simplify the encryption $Enc$ and decryption $Dec$ functions in our scheme to the following.
 
 %\begin{itemize}
\fakeparagraph{Enc'}: on input $m \in [0, M-1]$, samples $t \in \{0,1\}^n$  uniformly at random and encrypts message $m$ as $c := \langle m + F(k, t) \bmod M, t\rangle$.
\fakeparagraph{Dec'}: on input $c'$, decrypts ciphertext $c'= \langle c, t\rangle$  as $m = c - F(k, t) \bmod M$.
%\end{itemize}

\fakeparagraph{Claim 2.} \textit{If F is a pseudorandom function and $n=\lambda$ then the encryption scheme  $\Phi = (Gen,Enc',Dec')$ is CPA-secure.}

To proof the security of this construction, we first look at a hypothetical encryption scheme, which replaces the pseudorandom function $F_k$ with a randomly chosen function $R$ from the same domain. 
We argue with a proof by reduction that an attacker has only a negligible higher success probability in breaking the scheme with a pseudorandom function $F_k$ compared to a truly random function $R$. 
In a final step, we analyze an attacker for the scheme with a completely random function R. 

\textit{Proof.} Given scheme $\Phi = (Gen,Enc',Dec')$ we consider a second scheme $\tilde{\Phi} = (Gen*,Enc*,Dec*)$, which replaces the pseudorandom function $F_k$ with a random function $R$.
%\begin{itemize}
\fakeparagraph{Gen*}: on input $1^\lambda$, chose a uniform random function $R: \{0,1\}^n \rightarrow \{0,1\}^\lambda$ and set the plaintext space to $[0, M-1]$ where $M=2^\lambda$.
\fakeparagraph{Enc*}: on input $m \in [0,M-1]$, samples $t \in \{0,1\}^n$  uniformly at random and encrypts message $m$ as $c := \langle m + R(t) \bmod M, t\rangle$.
\fakeparagraph{Dec*}: on input $c'$, decrypts ciphertext $c'= \langle c, t\rangle$  as $m = c - R(t) \bmod M$.
%\end{itemize}

We first proof that a PPT adversary $A$ has only a negligible advantage in breaking scheme $\Phi$ compared to $\tilde{\Phi}$. 
We prove this by reduction assuming there exists a PPT adversary $A$ that has a non-negligible advantage in the CPA game with scheme $\Phi$ in comparison to $\tilde{\Phi}$. 
With the attacker $A$, we can construct a distinguisher $D$, which distinguishes a pseudorandom random function $F$ from a random function $R$ with non-negligible probability. 
In the reduction, $D$ has access to an oracle function $O: \{0,1\}^n \rightarrow \{0,1\}^\lambda$, and determines if this function is random or pseudorandom by emulating the attacker A.
%We construct the distinguisher $D$ as follows: 
Whenever $A$ performs a query to the encryption oracle with message $m \in [0, M-1]$, $D$ samples $r \in \{0,1\}^n$ uniformly at random, queries $Q(r)$ with response $x$ and responds to the attacker $A$ with $\langle m + x \bmod M, r\rangle$.
When $A$ gives two messages $m_0, m_1 \in [0, M-1]$ as an output, the distinguisher $D$ choses a bit $b \in \{0,1\}$ and samples $r \in \{0,1\}^n$ uniformly at random, queries $Q(r)$ with response $x$ and responds with $\langle m_b + x \bmod M, r\rangle$
When $A$ outputs the bit $b'$, the distinguisher D outputs 1 if $b'=b$ or 0 otherwise. 
We can make two observations given the distinguisher $D$.
If the oracle function $O$ is a pseudorandom function, the distinguisher $D$ has the same distribution as the attacker in the CPA experiment with the scheme $\Phi$. 
Similarly, if the oracle is a random function, the distinguisher $D$ has the same distribution as the attacker in the CPA experiment with the scheme $\tilde{\Phi}$.
Hence, given $F$ is a pseudorandom function, the probability advantage of adversary $A$ in succeeding in the CPA game with scheme $\Phi$ over the scheme $\tilde{\Phi}$ is the same success probability a distinguisher has in distinguishing a pseudorandom function from a random function. 

In the second part of the proof, we analyze the scheme $\tilde{\Phi}$ assuming a random function $R$. 
In the CPA game, the attacker $A$ can first query the encryption oracle $q(n)$ times before the challenge.
Assuming the challenge is computed as $\langle m + R(t') \bmod M, t'\rangle$, where $t'$ denotes the chosen random string, there are two possible outcomes.
In the first case, $t'$ never occurred as a choice in the querying phase. Since $R$ is truly random, $A$ has not learned anything in the querying phase.
As a result, the probability that $A$ outputs $b'=b$ during the challenge is $1/2$ because the output $R(t')$ is uniformly distributed and independent. 
The resulting ciphertext $c$ is the addition of  $R(t')$ and $m_b$ modulo $M$. Since $Prob[R(t') + m = c] = Prob[R(t') = c - m] = Prob[R(t')=r']$ and $r'$ is uniformly distributed, we can directly see that the probability is $1/2$ for distinguishing two encrypted messages.

If the attacker $A$ observes $t'$ in the querying phase, the attacker can determine which message was encrypted. 
Due to the observation of the ciphertext $c' = \langle m + R(t') \bmod M, t'\rangle$, the attacker learns that $m-c'=R(t')$, which may be used to distinguish the encrypted message.
The probability that the attacker observes $t'$ is smaller than $q(n)/2^{n}$, if $t'$ is uniformly drawn from $\{0,1\}^n$ and the number of queries of the attacker is bounded by the polynomial function $q(n)$. 

By combining the results from both cases and assuming $n = \lambda$, we can bound the success probability of the attacker $A$ in the CPA game with scheme $\tilde{\Phi}$ by the probability $1/2 + q(n)/2^n$.

Using the findings from the first part of the proof, we can bound the probability of the attacker $A$ in the CPA game with scheme $\Phi$ by the probability $1/2 + q(n)/2^n + \epsilon(n)$, where $\epsilon$ is a negligible function.
Because $q(n)/2^n$ is a negligible function and the addition of two negligible functions is again negligible, we complete the proof.

\subsubsection{Discussion}

\fakeparagraph{Length-Matching Hash Function.}
In our proof, we assume that the plaintext space matches the security parameter $\lambda$ (i.e., $M=2^\lambda$). 
However, if we only want to encrypt 64-bit integers with 128-bit outputs, we would have an overhead in the ciphertext size of 64-bits per ciphertext. 
To match the output of a PRF to the desired bits of $M$, one could use a length-matching hash function, which is analyzed in the Castelluccia encryption scheme~\cite{castelluccia2009}.
%A length preserving hash function $h : \{0,1\}^\lambda \rightarrow \{0,1\}^l$ must have the property that if $t$ is uniformly distributed over $\{0,1\}^\lambda$ then $h(t)$ should be uniformly distributed over $\{0,1\}^l$.
%Note that $h$ is not a cryptographic hash function (i.e., no collision resistance is required). 
%One possible construction of $h$ is splitting the output of the PRF into substrings of the desired range and exclusive-OR them together~\cite{castelluccia2009}.
%Since $h$ again outputs uniformly distributed strings with smaller length, the security proof only needs a few modifications if $h$ is applied to each output of the PRF.

%\todo{Length-Matching Hash Function, Sharing property of the tree proofs}
\fakeparagraph{Selection of the Key Identifier.} 
Our formal construction samples the identifier $t$, which serves as an input for the PRF, uniformly at random from $\{0,1\}^n$, to prove the CPA-security.
In our system,  $t$ is the identifier for the key being derived and represents a time counter.
As long as each identifier is only used once in the encryption process per stream (we keep the state on the client side), the scheme remains secure. 
Furthermore, the total number of keys can be selected according to the upper bound of chunks to be encrypted for a stream.

\subsection{Integrity with HoMAC}
\lsec{proof:integrity}
\oursystem employs the Homomorphic Message Authenticators scheme (HoMAC) presented in~\cite{catalano2013practical}.
For the poof of security, we refer to proof in~\cite{catalano2013practical}.
Instead of a standard PRF, \oursystem uses the key-canceling function to derive the nonces for each label. 
In \rsec{secproof}, we show that the key-canceling function is a PRF if a PRF is used to generate the canceling nonces.
In the proof they show that if the used function is a PRF, then the homomorphic MAC scheme is secure against the defined adversary. 

\subsection{Access Control Collusion}
\lsec{ac:leakage}
The constraint PRF TreeKD enables to share keys for intervals efficiently.
In the following, we discuss a performance trade-off with access control sharing multiple intervals of the same stream. 
\fakeparagraph{On-Off Shared Intervals.}
The encoding for the range-aggregation requires only two key derivations for in-range aggregated ciphertexts since the inner keys are canceled out. 
However, the encoding has a limitation.
If a stream owner shares two distinct time intervals $[t_0, t_2)$ and $[t_7, t_9)$ with the same user, the user is also able to compute the aggregation of the interval $[t_2, t_7)$, which lies between the two shared intervals.
This is due to the fact that inner keys cancel out.
The aggregation of the \chunksum for time interval $[t_i, t_j)$ is encrypted as $\sum_{x=i}^{j-1}  c_{x} = \sum_{x=i}^{j-1}  m_{x} + k_i - k_j$.
With access to the intervals $[t_0, t_2)$ and $[t_7, t_9)$, the data consumer can derive the keys $\{k_0, k_1, k_2, k_7, k_8, k_9\}$. 
Hence, in addition to the ciphertexts within the interval, the data consumer can also decrypt the aggregated ciphertext for the interval $[t_2, t_7)$ as $\sum_{x=2}^{6} m_{x} = \sum_{x=2}^{6} c_{x} - k_2 + k_7$.
The ramifications of this issue arise when users share adjacent intervals in the same stream with small gaps. 
\oursystem provides a hybrid key-canceling mechanism that limits this leakage in a trade-off for longer decryption times.
We split the keys into epochs by replacing some $k_i$ with non-cancelling skip-keys $k'_i, k''_i$ in $k_{i-1} - k_i$ and $k_i - k_{i+1}$, respectively. 
With this, we can share one interval per epoch without leakage. 
This increases the cost of aggregations over the epoch borders by two key derivations and an addition.

\end{document}